\documentclass[acmsmall]{acmart} 

\AtBeginDocument{%
  }

\usepackage{tabularx}
\usepackage{float} 
\usepackage{graphicx} 
\usepackage{booktabs} 

\begin{document}

\title{Black Older Adults’ Perception of Using Voice Assistants to Enact a Medical Recovery Curriculum}

\author{Andrea Green}
\orcid{0009-0004-6973-4552}
\affiliation{%
    \department{Civil and Environmental Engineering}
  \institution{Stanford University}
  \city{Stanford}
  \state{CA}
  \country{USA}}
  \email{andgreen@stanford.edu}

\author{Gabrielle Polite}
\orcid{0009-0006-4065-2216}
\affiliation{%
    \department{Computer Science}
  \institution{Stanford University}
  \city{Stanford}
  \state{CA}
  \country{USA}}
  \email{gpolite@stanford.edu}

\author{Isabelle Hung}
\orcid{0009-0006-2952-8580}
\affiliation{%
  \institution{Parsons School of Design}
  \city{New York}
  \state{NY}
  \country{USA}}

\author{Kristen L. Fessele}
\orcid{0000-0003-2336-6841}
\affiliation{
   \institution{Memorial Sloan Kettering Cancer Center}
   \city{New York}
   \state{NY}
   \country{USA}}
   \email{fesselek@mskcc.org}

\author{Sarah L. Billington}
\orcid{0000-0003-3796-587X}
\affiliation{%
    \department{Civil and Environmental Engineering}
  \institution{Stanford University}
  \city{Stanford}
  \state{CA}
  \country{USA}}
\email{billingt@stanford.edu}

\author{James A. Landay}
\orcid{0000-0003-1520-8894}
\affiliation{%
    \department{Computer Science}
  \institution{Stanford University, Institute for Human-Centered Artificial Intelligence}
  \city{Stanford}
  \state{CA}
  \country{USA}}
\email{landay@stanford.edu}

\author{Andrea Cuadra}
\orcid{0000-0002-1845-0240}
\affiliation{%
  \institution{Olin College of Engineering}
  \city{Needham}
  \state{MA}
  \country{USA}}
\email{acuadra@olin.edu}

\renewcommand{\shortauthors}{Andrea Green et al.}

\begin{abstract}
The use of interactive voice assistants (IVAs) in healthcare provides an avenue to address diverse health needs, such as gaps in the medical recovery period for older adult patients who have recently experienced serious illness. By using a voice-assisted medical recovery curriculum, discharged patients can receive ongoing support as they recover. However, there exist significant medical and technology disparities among older adults, particularly among Black older adults. We recruited 26 Black older adults to participate in the design process of an IVA-enacted medical recovery curriculum by providing feedback during the early stages of design. Lack of cultural relevancy, accountability, privacy concerns, and stigmas associated with aging and disability made participants reluctant to engage with the technology unless in a position of extreme need. This study underscored the need for Black cultural representation, whether it regarded the IVA's accent, the types of media featured, or race-specific medical advice, and the need for strategies to address participants' concerns and stigmas. Participants saw the value in the curriculum for those who did not have caregivers and deliberated about the trade-offs the technology presented. We discuss tensions surrounding inclusion and representation and conclude by showing how we enacted the lessons from this study in future design plans. 
\end{abstract}



\setcopyright{acmlicensed}
\acmJournal{PACMHCI}
\acmYear{2025} \acmVolume{9} \acmNumber{2} \acmArticle{CSCW039} \acmMonth{4}\acmDOI{10.1145/3710937}

\setcopyright{none}
\copyrightyear{2025}
\acmYear{2025}
\acmConference[CSCW '25]{The 28th ACM SIGCHI Conference on Computer-Supported Cooperative Work \& Social Computing}{October 18-22, 2025}{Bergen, Norway}
\acmDOI{10.1145/3710937}

\makeatletter
  \@printpermissiontrue
  \@printcopyrighttrue
  
  \renewcommand\@copyrightpermission{%
    \footnotesize
    This is the authors' version of the work. It is posted here for your personal use, not for redistribution. The definitive Version of Record was published in:
    \par
  }

  \renewcommand\@copyrightowner{%
  Copyright held by the owner/author(s).
  }
\makeatother
\received{January 2024}
\received[revised]{July 2024}
\received[accepted]{October 2024}

\maketitle

\section{Introduction}
In an ideal world, we would all have the care we need at the times we need it. However, in reality, health services are under-resourced \cite{accessHealthServices, underservedPops}, specifically for racial minorities (e.g., American Indians, African Americans, and Latin Americans). Additionally, healthcare is unequally distributed \cite{underservedPops, gibbons2010reducing, millery2010health, villarosa2022under}, exacerbating disparities. One of the key challenges in healthcare is providing personalized care at scale. Personalized care is a patient-centered approach that aims to tailor treatment and services to the unique needs of each individual patient. 
Older adult patients, who frequently have multiple concurrent health issues, especially stand to benefit from personalized care that considers their physical, cognitive, and social needs. Personalized care can also play a crucial role in addressing healthcare disparities for underserved and marginalized groups, such as racial and ethnic minorities, by accounting for factors like language, culture, and socioeconomic status. 

However, personalized care can be prohibitively expensive \cite{rose2013personalized, aronson2015making}, creating an urgent need for scalable and affordable innovations to fill the gaps in care. One example of such a gap occurs during the transition from intensive treatment for a life-threatening disease (e.g., cancer) to early survivorship. This period can be particularly challenging for individuals experiencing frailty \cite{muscedere2017impact,bagshaw2014association}, a condition more common for older adults \cite{mitnitski2016rate}. In this period, recently discharged older patients go from receiving a large amount of medical attention to almost no medical attention while in need of support acclimating to a new normal \cite{forster2003incidence}. Health promotion interventions such as incremental increases in overall physical activity, strength and balance training, and nutritional support may slow or reverse the frailty trajectory \cite{de2015effects, lang2009frailty, gill2011relationship}. However, there is a shortage of human resources to support this kind of home health intervention \cite{brown2023nurse, travers2023environmental, mukamel2023association}.

A promising opportunity to fill this gap is to pair the ubiquity of multimodal computing interfaces, such as smart speakers that have screens, with technological advances in conversational technologies, such as large language models (LLMs) \cite{chang2023survey} and models that reason over diverse input modality signals \cite{moon2023anymal}, to deliver health promotion interventions. The use of interactive voice assistants (IVAs) to interact with technology may be more intuitive and inclusive for older adults, especially for those with high degrees of medical frailty, lower digital literacy, or cognitive impairment \cite{pradhan2018accessibility, zubatiy2021empowering}. Intentionally designed with the needs of older adults at the forefront of the process, a multimodal, artificial intelligence-based interface could deliver personalized health promotion education to patients in their homes immediately after discharge. Such an intervention could significantly affect the course of medical recovery and increase a patient’s quality of life. 

However, technology \cite{toyama2015geek, buolamwini2018gender, buolamwini2017gender, benjamin2023race, eubanks2018automating} and healthcare \cite{williams2009discrimination, kung2008deaths, forsat2020recruitment} have been documented to exacerbate inequities. As promising as such home health technologies might sound, if the needs of those who stand to benefit from their promises are not carefully considered, they could create more harm than good.

Older adults \cite{chu2022digital} and the Black community \cite{rankin2021resisting} have been historically overlooked in the design of technology, which could make it less likely that they would feel comfortable adopting yet another new technology. Simultaneously, digital health interventions are expanding rapidly with little knowledge about acceptance and engagement by these groups \cite{krukowski2024digital}. Therefore, studies like this one are crucial for understanding the impact of digital interventions on Black older adults and the intersection of technology and medical recovery. 

To create more equitable technologies for filling gaps in health services, we conduct participatory design with Black older adults through five focus groups ($N$=26), and follow-up speed-dating sessions ($n$=15). In all study sessions, we discuss speculative designs for enacting the \textit{Care Curriculum}. In the focus groups, we present three concept videos of different ways to deliver physical activity education, one of the modules of care, to elicit feedback on novel ways of enacting medical recovery lessons. In the speed dating sessions—one-on-one sessions with individual participants rapidly reacting to many different ideas \cite{davidoff2007rapidly}---we present twelve storyboards. Each of these storyboards map to a different module of the \textit{Care Curriculum} we are interested in building (e.g., getting to know the device, nutrition, and breathing exercises). We address three main research questions:

\begin{enumerate}
    \item[\textbf{RQ1:}] \textbf{Motivation.} What motivates Black older adults to engage with the \textit{Care Curriculum}? 
    \item[\textbf{RQ2:}] \textbf{Perception of the technology.} How do Black older adults perceive using IVAs to enact the \textit{Care Curriculum} in their homes?
    \item[\textbf{RQ3:}] \textbf{Opportunities and challenges.} What opportunities and challenges do they anticipate?
\end{enumerate}

In this study, we refer to the medical recovery curriculum as the \textit{Care Curriculum}. This paper’s contribution is a characterization of considerations we must take when creating health IVAs that are inclusive of Black older adults. While we found that Black representation is crucial for motivating our participants, appropriately representing all marginalized groups is a wicked problem \cite{rittel1973dilemmas} for many reasons described in the paper. Our participants had varied perceptions of IVAs, viewing them as companions or not, with positive reception for adaptability to different abilities. Anticipated opportunities involve leveraging user curiosity about machine learning to increase motivation to engage with the \textit{Care Curriculum}. Challenges include ensuring the appropriateness of the \textit{Care Curriculum} in terms of cultural nuances, increasing IVA accountability, and incorporating diverse accents. As a whole, this paper provides knowledge on how to design more equitable home health interfaces.

\section{Related Work}
In this section, we will describe research related to participatory design for home health interfaces in computer-supported cooperative work (CSCW), medical recovery care and racial disparities, IVAs for older adults, and IVAs for Black older adults.

\subsection{Participatory design for home health interfaces in CSCW}
Computer-Supported Cooperative Work (CSCW) as a field has long concerned itself with exploring human-computer interaction problems using a socio-technical lens. A central aim is to represent and address the needs of marginalized groups \cite{harrington2019deconstructing, roe2023examining}, including the Black \cite{twyman2017black}, fat\footnote{The term ``fat'' is used following the terminology recommendations (e.g., avoiding medical terminology) outlined by \citet{payne2023howtoethically} and is aligned with the concept of using this term for social justice, as discussed by \citet{mcphail2021fat}.}\cite{payne2023howtoethically}, and older adult \cite{choi2023together} communities. One recent example of how to include diverse voices is a method Haghighi \& Jörke et al.~\cite{haghighi2023workshop} developed to enable people with varying levels and areas of domain expertise and with a variety of lived experiences to collectively speculate about the ethical implications of emerging technologies, navigate value tensions, and prototype artifacts as a way to grapple with those tensions.

Going further, our paper aligns with \citet{harrington2019deconstructing}'s stance, which argues that participatory design ``is a privileged, white, youthful, and upper to middle-class approach to innovation that consists of activities that implore participants to rely on ideals of imagination, creativity, and novel insight,'' and there is a need to ``re-center the focus of design on individuals who are historically underserved.'' In our research, we investigate how to use ambient intelligence to bridge a gap between healthcare and technology with a focus on Black older adults, a group that is intersectionally marginalized and severely underrepresented in healthcare technical innovation. The technology we are speculating about in this work, the \textit{Care Curriculum}, might inevitably affect all communities at some point, and we aim to investigate the needs of Black older adults before the technology is built. As a result, designers, medical researchers, and developers will have the necessary resources available to represent Black people. 

\subsection{Medical recovery and racial disparities}
Medical recovery care, also known as post-hospitalization care, is the medical support that patients receive after they have been discharged from the hospital following an illness, injury, or surgical procedure \cite{RisingCare,Bairwa}. This support may include home nurses to monitor vital signs, personal care assistance with activities of daily living, dietary and nutrition management, and emotion counseling \cite{RisingCare,Bairwa}. This care has been shown to benefit the patient and the healthcare system by reducing the rate of re-injury and additional hospitalizations \cite{Bairwa,misky2010post}. 

The increasing demand for healthcare services has led to the expansion of IVAs in the healthcare industry  \cite{sezgin2020readiness}. Research focused on caregivers has found that IVAs can be used to fill gaps in access to information and support, such as by updating or personalizing a patient’s care plan or providing new ideas \cite{bartle2022second}, or to support complex home care \cite{tennant2022caregiver, bartle2023machine}.

Although recovery care is a benefit, prior research has highlighted racial disparities in this type of care for Black patients \cite{fat2019racial,englum2011racial, odonkor2021disparities}. Black patients are less likely to receive or use recovery treatments after leaving the hospital \cite{fat2019racial, englum2011racial, de2007ethnic}. Instead, they are more frequently instructed to return home rather than being referred to facilities for recovery care \cite{englum2011racial,de2007ethnic}, despite potentially having more post-operative complications \cite{wood2019association}. This disparity in treatment recommendations could be related to unequal access to recovery services \cite{englum2011racial,odonkor2021disparities} or the challenges of being uninsured or under-insured through Medicaid \cite{shen2016racial,shen2007disparities,shen2001exploration,burstin1992socioeconomic} among racial and ethnic groups \cite{sohn2017racial}.  

Another factor contributing to recovery disparities and access challenges may be rooted in racism within the healthcare system, particularly concerning the treatment of pain in Black patients in recovery. Research has shown that Black patients often receive less pain medication during recovery compared to white patients \cite{harbell2023addressing, anderson2009racial, aronowitz2020mixed}. This disparity stems from a misconception among some healthcare providers who believe that Black patients have a higher pain tolerance than white patients \cite{trawalter2012racial, anastas2024impact}. 

These studies highlight the importance of conducting additional research and implementing interventions to address racial disparities in access to medical recovery care. Our paper expands this work to a \textit{Care Curriculum} designed to address gaps in ongoing treatment for patients recovering from a serious illness. 
Including the perspective of Black older adults in this research mitigates the lack of representation that contributes to existing racial disparities caused by limited access to recovery care. 

\subsection{IVAs for older adults}
Research on older adults' use of IVAs has revealed several key insights, finding that they primarily use these devices for reminders and obtaining information \cite{arnold2022does,o2020voice,islam2022framework}, companionship and entertainment \cite{o2020voice}, and obtaining basic health information \cite{brewer2022empirical, arnold2022does}. Our study builds on this foundation by explicitly examining how older adults integrate IVAs into recovery care, thereby predetermining the functionality within this community. Therefore, our contribution lies in identifying factors that support older adults recovering after a medical discharge using novel interaction mechanisms with IVAs. 

Beyond functionality, studies have explored older adults' perceptions of IVAs, uncovering a complex relationship \cite{spangler2022privacy,thankachan2023challenges,liu2023older}. While some older adults appreciate the user-friendliness and communication experience of IVAs, viewing them as potential companions for emotional support and enjoyment \cite{kim2021exploring, liu2023older}, others express skepticism \cite{cuadra2023designing, upadhyay2021empirical}, mistrust \cite{upadhyay2021empirical}, and even concerns about the IVA manipulating them \cite{horstmann2023alexa}. Another study found that older adults often perceive IVAs as both human-like and object-like, shifting between these categorizations during interactions \cite{pradhan2019phantom}. These varied findings in existing literature motivate our research to specifically investigate Black older adults' perceptions of IVAs, aiming to understand how their perspectives may align with or differ from the broader older adult community. 

\subsection{IVAs for Black older adults} 
Research on Black older adults' perceptions of IVAs has found that Black older adults often feel their culture is not adequately represented in the voices and interactions of IVAs \cite{brewer2023envisioning, harrington_its_2022} and has identified specific communication challenges that Black older adults face when interacting with them \cite{harrington_its_2022}. These challenges may include difficulties with voice recognition, misunderstandings of cultural references or expressions, and the need for users to adjust their natural speech patterns to be understood by the IVA \cite{harrington_its_2022}.  Researchers emphasize the importance of the authentic inclusion of Black voices, raising questions about the capability of LLMs to achieve this authenticity \cite{brewer2023envisioning}. Our study seeks to expand on the literature by investigating how IVAs might adapt to improve communication effectiveness and suit the specific requirements of Black older adults in the context of the \textit{Care Curriculum}. 

A study by \citet{harrington_its_2022} found that Black culture plays a role in shaping Black older adults’ perceptions of IVAs which is a similar finding to that of \citet{cuadra2023designing}'s research on designing voice-first ambient interfaces for older adults. The study suggests that the perception of IVAs among older adults is influenced by lived experiences. \citet{pradhan2019phantom} recommend that future studies focus more on understanding how cultural factors influence older adults’ perceptions of IVAs. 

Some studies on older adults’ perception of IVAs have not disclosed the racial background of the participants, which can lead to ambiguity when interpreting whether the findings are generalizable to the Black older adult community. Our study seeks to fill a gap in the literature by studying how Black older adults perceive medical recovery care delivered by an IVA. Given the wide variety of racial disparities in healthcare and HCI, a CSCW paper by \citet{roe2023examining} suggests that the explicit discussion and analysis of race should be pivotal to examining racial disparities and race as a category of human difference in greater depth. In our study, we explicitly recruit older adults who self-identify as Black and have lived in the United States for most of their lives. Given how difficult it is to define race \cite{ogbonnaya2020critical}, in this paper, we avoid choosing one specific definition and instead refer to Blackness as individuals who belong to the Black diaspora, align with Black cultures, and self-identify as Black. 

\section{Method}
We conducted a human participant study using a combination of video elicitation focus groups and one-on-one speed dating sessions \cite{zimmerman2017speed,bartle2022second} to gather data about how to represent the needs and preferences of Black older adults in the design of the \textit{Care Curriculum}. This section will describe our participants, the \textit{Care Curriculum}, our design probes and how they complemented one another, procedures, analytical approach, and researcher positionality.  

\subsection{Participants}
We recruited Black older adult participants ($N$=26; 14 women and 12 men) ages 59+ ($\overline{M}$=69.7, $\tilde{SD}$=5.41) using snowball sampling via flyers, email, and word of mouth. The research followed ethical standards and received approval from our institutional review board (IRB \#48481). 
Our inclusion criteria were that all participants had to be Black-identifying older adults aged 55+.  
For each focus group, participants knew each other through affinity groups (e.g., a sewing group and a college fraternity). All participants had lived in the US for over 30 years, were not of Hispanic origin, and reported speaking English as their primary language with no reports of a second language. Participants reported the following education levels: high-school diploma/GED ($n$=1), Associate's degree ($n$=1), Bachelor's degree ($n$=6), Master’s degree ($n$=13), and Juris doctorate ($n$=1). They had previous employment experience with wages or salary in sectors of administration and management ($n$=5), government and public service ($n$=5), education ($n$=4), finance and accounting ($n$=4), sales and marketing ($n$=4), engineering and project management ($n$=2), and housing ($n$=2). All participants were confident in their reading and writing abilities and had prior experience with computing devices. Most of the participants were born in the U.S. ($n$=24), were retired ($n$=19), married ($n$=14), and earned over \$75,000 annually ($n$=17). Most reported owning computing devices ($n$=25), used the internet multiple times every day ($n$=25), and accessed the internet at home via WiFi ($n$=25). Most were very confident using computing and speech-based computing devices ($n$=19). Some participants lived alone ($n$=9), but most lived with at least one other person ($n$=17). Table \ref{tab:demographics} contains participant demographic details. 

\subsubsection{Experience with medical recovery}
Participants with a range of medical recovery experiences were included. Most participants ($n$=17) explicitly reported having a variety of experience with recovery, including: their own past recovery experiences ($n$=10); ongoing recovery efforts due to complications from conditions such as high cholesterol, cancer, and diabetes ($n$=4); and caring for an older adult who was or had been in recovery ($n$=5). 

Our approach includes older adults regardless of their medical recovery experiences, broadening the range of perspectives in our study. This inclusive method enhances collaboration by distributing decision-making power more evenly among participants, rather than just focusing on the participants whose perspectives inform the final product \cite{harrington2019deconstructing}.

\begin{table}
\caption{Self-reported Participant Demographics ($N$=26)}
\centering
\begin{tabularx}{\textwidth}{lX}
\toprule
Age group & 55-59: 1, 60-64: 3, 65-69: 9, 70-74: 7, 75-79: 6 \\
\midrule
Gender & Woman: 14, Man: 12 \\
\midrule
People living in household &  One person: 9, Two people: 14, Three people: 2, Five or more people: 1\\
\midrule
Current marital status & Married: 14, Never married: 3, Living with a partner: 1, Divorced: 5, Widowed: 2, Not married: 1 \\
\midrule
Birth country &  USA: 24, Germany: 2 \\
\midrule
Highest degree of schooling & High-school diploma/GED: 1, Some college: 4, Associate's degree: 1, Bachelor's degree: 6, Master's degree: 13, Juris doctorate: 1 \\
\midrule
Employment background & Administration and Management: 5, Government and Public service: 5, Education: 4, Finance and Accounting: 4, Sales and Marketing: 4, Engineering and Project Management: 2, Housing: 2 \\
\midrule
Retired &  Yes: 19, No: 7 \\
\midrule
Yearly household income range & \$35,000--\$49,999: 2, \$50,000--\$74,999: 4, >\$75,000: 17, Unreported: 3 \\
\midrule
Computing device(s) owned & Smartphone: 22, Laptop computer: 19, Desktop computer: 18, Tablet: 20, Smart speakers: 10, Unreported: 1 \\
\midrule
Confidence with computing device & Somewhat not confident: 1, Somewhat confident: 2, Confident: 4, Very confident: 19 \\
\midrule
Confidence with speech-based device & Not confident: 2, Somewhat not confident: 4, Neither confident nor not confident: 4, Somewhat confident: 4, Confident: 12  \\
\bottomrule
\end{tabularx}
\newline
\label{tab:demographics}
\end{table}

\subsection{Care Curriculum design}
\label{sec:carecurriculumdesign}

The content of the \textit{Care Curriculum} used in this study was developed by a team of oncology specialists (i.e., a nurse scientist, a rehabilitation medicine physician, and an oncology nurse), from a medical institution we collaborated with, using cancer survivorship clinical practice guidelines from the American Cancer Society  \cite{rock2022american}, National Comprehensive Cancer Network \cite{national2023nccn}, American Society of Clinical Oncologists \cite{ligibel2022exercise}, and the World Cancer Research Fund \cite{shams2019operationalizing}. While those guidelines are written in the context of a cancer survivorship population that requires long-term follow-up for residual effects of treatment and disease recurrence, most of the recommendations reflect nutrition, physical activity, and healthy aging guidance appropriate for the general public \cite{nationalhealthinitiatives} as reflected in our storyboards (see Table \ref{tab:designprobes}). The curriculum begins with an overview of the evidence supporting these healthy lifestyle interventions to prevent or manage co-morbid conditions (e.g., heart disease and diabetes), maintain independent mobility and cognitive function, and resolve treatment effects such as fatigue. Our design probes are representations of the \textit{Care Curriculum}. 

We directly collaborated with a member of the oncology team, who also serves as one of the co-authors of this paper, to determine older adults' perception of the \textit{Care Curriculum} and the nature of using an IVA to deliver it to develop design implications. The curriculum was developed by the oncology specialists as a component of an upcoming larger study to reintroduce healthy lifestyle principles to older adult survivors of cancer. The larger study, which will be conducted by the specialists, will recruit older adults of all races who have completed their anticancer therapy within the past three months and who will interact with a web-based, holistic multimodal large language model-based interactive voice assistant to complete a geriatric assessment \cite{cuadra2024digital}. Tailored to the assessment responses, the system will then deliver weekly modules of educational content adapted from national cancer survivorship clinical practice guidelines. For example, a participant who reports spending most of their day in bed due to pain will initially receive bed exercises to safely strengthen the lower extremities (as seen in the \textit{Physical Activity} storyboard), whereas a fully ambulatory participant might receive a walking program and standing balance exercises (as seen in the \textit{Balance} storyboard). 

\subsection{Design probe 1: Concept videos for focus groups}

\begin{figure}
    \includegraphics[width=\textwidth]{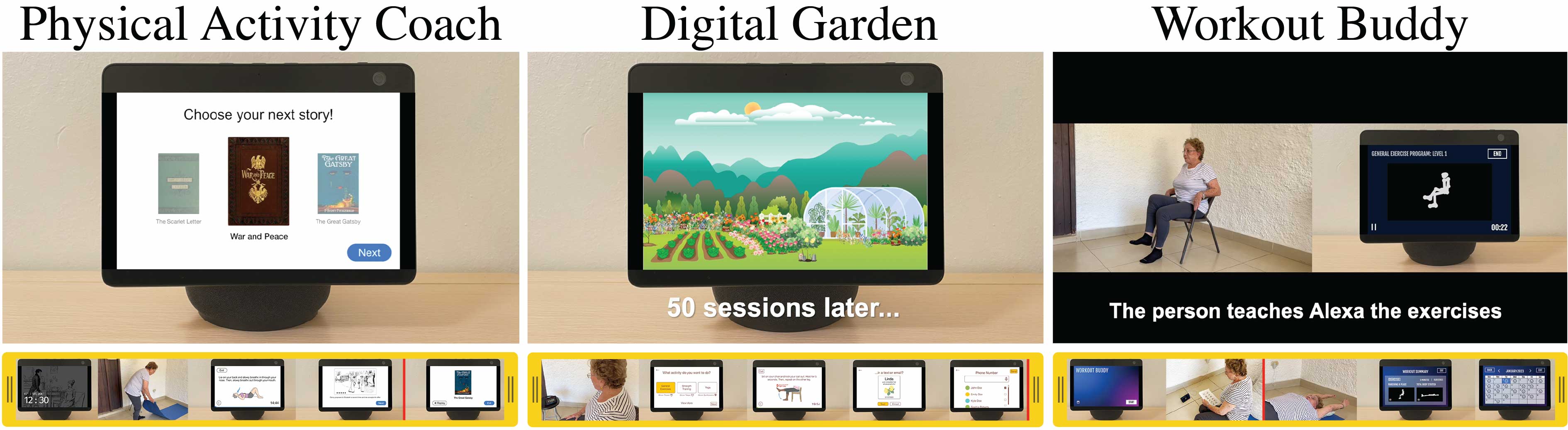}
    \caption{Screenshots from each concept video: \textit{Physical Activity Coach}, \textit{Digital Garden}, and \textit{Workout Buddy}. Note, that the IVA in the videos is called Alexa for participants to make the connection to a mainstream IVA. The IVA in the storyboard is called Carol, due to its role of ``caring'' for others and to further convey that it is an imaginary idea.    
    }\label{fig:conceptvideos}
\end{figure}

We used three concept videos that showed an older adult interacting with the IVA in different scenarios to encourage physical activity (see Figure \ref{fig:conceptvideos} and Table \ref{tab:designprobes}). The first video, \textit{Physical Activity Coach}, showed the user, an older woman, using the IVA to complete a series of breathing exercises either while listening to or while the IVA reads an audiobook aloud to her. The second video, \textit{Digital Garden}, showed the user building a digital garden---every time she completed an exercise, a new flower was planted, and she could share her success with friends and family. The third video, \textit{Workout Buddy}, showed the user teaching the IVA how to perform a set of exercises she received as a printout. The person depicted using the technology in the videos is an older Latinx woman. Note, despite creating the concept videos with inclusion in mind, we did not create them specifically tailored to Black patients. We chose to present them to older Black adults to understand more deeply how the curriculum would be received by a group we know has been disproportionately excluded in the design of medical technologies. Detailed descriptions of each video are as follows:

\begin{table}[h!]
\caption{Summary of the Design Probes}
\centering
\begin{tabularx}{\textwidth}{lX}
\toprule
\textbf{Concept Video} \newline &  \textbf{Description} \\
\midrule
1. Physical Activity Coach & The VA leads an older adult through a series of breathing exercises while reading a book to her. \\
\midrule
2. Digital Garden & The VA helps the older adult through a series of exercises; each time she completes an exercise, a flower is added to her garden. \\
\midrule
3. Workout Buddy & The older adult shows the VA the exercises she learned at physical therapy; the VA remembers these exercises and can recall them at a later date. \\
\bottomrule
\textbf{Storyboard}  \newline &  \textbf{Description} \\
\midrule
1. Getting to Know Carol & The VA (Carol) introduces Nana to the program and walks her through orientation videos. \\
\midrule
2. Goal Setting & Carol helps Nana create a goal of stretching for 10 min/day, and reminds her of it the following day. \\
\midrule
3. Physical Activity & Carol helps Nana perform bed exercises since her back has been in pain, which leads Nana to feeling better mentally and physically because she still completed her goal. \\
\midrule
4. Staying Safe \newline  & Carol shows Nana a video about how to avoid falls. \\
\midrule
5. Feeling Fatigued & Carol leads Nana through a dance exercise because Nana was tired but she still wanted to exercise. \\
\midrule
6. Balance & Carol notices that Nana has been struggling with her balance, so Carol asks Nana to perform some balance exercises.  \\
\midrule
7. Working Out & Carol sees that Nana has been getting stronger, so Carol asks her to do a more advanced workout. \\
\midrule
8. Nutrition & Carol takes account of the foods in Nana's kitchen and creates a recipe that includes all the major food groups, which Nana really appreciates and values. \\
\midrule
9. Healthy Weight & Carol notices that Nana has gained 20 lbs in the last month, so she asks Nana about her symptoms recently. Carol decides to call Nana's doctor to set up an appointment. \\
\midrule
10. Healthy Mind & Carol asks Nana to perform a set of breathing exercises since Nana has been stressed recently. \\
\midrule
11. Positive Reminiscing & Carol asks Nana to think about her favorite memory and then recount those stories to Carol which really boosts Nana's mood.  \\
\midrule
12. Progress Check  \newline  & Carol asks Nana a series of questions about her experience in the program thus far. \\
\bottomrule
\end{tabularx}
\label{tab:designprobes}
\end{table}

\paragraph{Physical Activity Coach} The IVA leads the user through a 15-minute, moderate intensity breathing exercise. As the user is completing the exercise, the IVA is reading excerpts of \textit{Pride and Prejudice} by Jane Austen. The integration of the narrative element is inspired by prior HCI research findings that narrative-based interfaces are effective tools for motivating physical activity \cite{murnane2020designing, murnane2023narrative}. The IVA asks the user to get a yoga mat and a chair to help with the exercises. The user completes the set of breathing exercises at the direction of the IVA. For example, the IVA says, ``\textit{Lie on your back and slowly breathe in through your nose. Then, slowly breathe out through your mouth.}'' The IVA continues to read the book excerpts in between the instructions. After the 15 minutes are over, the IVA provides three book options to read next: \textit{The Scarlet Letter} by Nathaniel Hawthorne, \textit{War and Peace} by Leo Tolstoy, and \textit{The Great Gatsby} by F. Scott Fitzgerald. All books were randomly selected from Project Gutenberg's list of free eBooks. 

\paragraph{Digital Garden} The IVA displays three activities: general exercises, strength training, and yoga. Each exercise corresponds to a different type of flower that will be put in their digital garden afterwards. The garden metaphor is an adaptation of prior work in HCI effectively using garden metaphors to motivate behavior change \cite{consolvo2008activity, froehlich2009ubigreen}. The user says ``general exercises,'' and then selects \textit{20 minutes} for the length and \textit{light} for the intensity of their workout. After the IVA leads the user through a series of exercises, including leg raises, the user has the option to share their accomplishment with one of their contacts. The user then selects a contact from their contact list and sends him a message stating her accomplishment. Afterward, a new flower is added to the user's virtual garden. After 50 sessions, the user's garden is full of flowers. 

\paragraph{Workout Buddy} The user uses the IVA as a workout buddy. The user had previously received a list of exercises to perform from their physical therapist, and ``teaches'' these exercises to the IVA. This flip from framing the older adult as a recipient of information, to instead being the provider of that information was inspired by the experience of a participant reported in prior work \cite{cuadra2023designing} who wanted to teach the IVA, an Amazon Alexa, to be smarter. The IVA takes note of these exercises and records them to keep in the repertoire of possible movements. After the user completes the exercises assigned to them, the video cuts to a few weeks later when the user asks the IVA to recall \textit{``What was the exercise where I was sitting on a chair and moving my leg up?}'' The IVA responds by stating the date this exercise was performed and showing the user a video to jog their memory of how to complete the exercise. 

\subsection{Design probe 2: Storyboards for speed dating sessions}
We created 12 storyboards, each based on modules from a 12-week recovery plan (see in Table \ref{tab:designprobes}) as described in Section \ref{sec:carecurriculumdesign}. 
Each storyboard focused on a specific activity, including stretching, positive reminiscing, nutrition, and healthy weight (see Figure \ref{fig:storyboards} for example). The storyboards were introduced to provide participants with an overview of what the recovery plan includes. The physical form of Carol, the IVA, was intentionally left ambiguous in the storyboards, allowing viewers to focus more on the medical scenarios being depicted and the interactions between Carol and Nana, the older woman she was assisting, rather than the specific details of Carol's underlying technology. The storyboards were designed in a simple, minimalistic visual style, avoiding overly technical or detailed representations of Carol, with a text description to depict the functions and emotions of the scene. This design and use of storyboards is an effective HCI technique \cite{bartle2022second} because it allows participants to laterally imagine alternatives. 

\begin{figure}
\label{fig:storyboards}
\includegraphics[width=\textwidth]{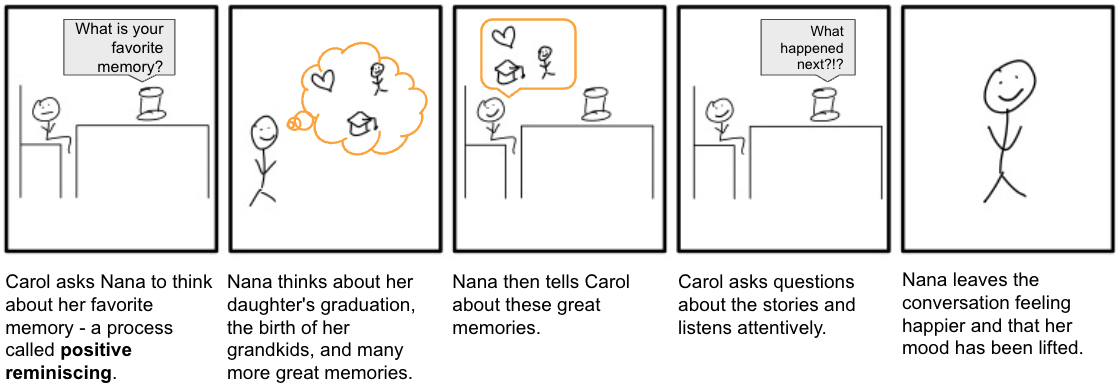}
    \Description[This is a short title for what's on the image.]{This is a long description for accessibility.}
    \caption{Storyboard called \textit{Positive Reminiscing}.}\label{fig:storyboards}
\end{figure}

\subsection{Utilizing complementary design probes}
We used high- and low-fidelity design probes, complementary, to gain insights into different aspects of the \textit{Care Curriculum.} 
\textbf{High-fidelity prototyping}, such as concept videos, allow participants to see the interaction concretely and see the context of how it might fit into their lives as they ``offer more realistic interactions and are better at conveying the range of design possibilities'' \cite{walker2002high}. \textbf{Low-fidelity prototyping}, such as storyboards, provide details of different scenarios and new ideas of lateral scenarios based on each individual's own lived experiences as they allow designers and users the ability to ``focus on high-level interaction design and information architecture, rather than on details or visual style'' \cite{walker2002high}. 

Moreover, the focus group and individual one-on-one participant configurations were also carefully selected. For group discussions, we selected videos that showcased physical activity because it is rarely stigmatized, creating the space for participants to comment on the implementation methods. The participants in these groups knew each other through community-based activities, so it was important to understand the general perceptions of the group towards the technologies we were presenting. Storyboards, which included more-often-stigmatized \textit{Care Curriculum} topics, such as weight management and mental health, were used in individual settings where the conversations were more private. Our aim was to obtain more feedback on the content of the specific curriculum modules, allowing  for a deeper, more personal exploration of these sensitive issues. Since the concept videos were presented in a group setting, there was a small risk of carry-over effects introduced. However, participants focused on different aspects of the research, as planned, during the storyboard sessions. 

The combination of methods employed allowed participants to ground their observations on the concrete interactions demonstrated in the concept videos without getting off-topic by high-fidelity elements, such as the race or accent of the people in the videos, during the storyboard sessions. However, a few participants encountered difficulties with certain storyboard scenarios because the details of how the technology functioned were unclear. This created a minor barrier to their participation.

\subsection{Procedure}
Eligible participants who demonstrated interest were sent a digital consent form and asked whether they would like to participate in the focus groups only or also in the speed dating sessions. A researcher was virtually available to answer any questions they had about the study or the consent form. Once signed consent forms had been received, we scheduled the respective study sessions. All (except for one) speed dating session participants attended the focus groups before their speed dating sessions. There were five focus groups---\textbf{F1:} P1--P6, \textbf{F2:} P7--P11, \textbf{F3:} P12--P15, \textbf{F4:} P16--P20, and \textbf{F5:} P21--P26. The subset of focus group participants ($n$=15) that participated in the speed dating sessions were: P1, P3--P4, P7--P9, P12, P14--P17, P21, P23, and P25--P26. The speed dating sessions occurred within 3--7 days of the focus groups. Sessions were organized and carried out by the same Black researcher. 

\subsubsection{Focus groups}
Focus groups were carried out on Zoom. To start, all participants were given an introduction to the research project, including the following script, \textit{``We are investigating how voice-based technologies could be used to help older adults recovering from medical treatment. Imagine you had medical treatment, and you’re now at home recovering. We are going to show you some short videos and ask for your thoughts, impressions, and feedback. Please be honest! All of these are just ideas meant to generate more ideas, so feel free to be `mean' about these prototypes—the more honest feedback we get right now in the early stages of this work, the better the outcome will be in the long run.''}
Then, the participants watched each video and were asked a series of questions from the moderator to elicit their thoughts (e.g., ``Would you use something like this? If so, when do you imagine yourself using this?'', and ``What would you change about the voice assistant in the video?'') to foster discussions about the system, leveraging the additional details provided in the videos.
The moderator followed a structured guide and ensured that all participants had the opportunity to speak and provide their input on each video. Participants were also asked if they wanted to add any follow-up comments at the end, and some did choose to do so. The focus groups lasted an average of 63 minutes (\textbf{F1}: 60, \textbf{F2}: 62, \textbf{F3}: 70, \textbf{F4}: 54, \textbf{F5}: 69).

\subsubsection{Speed dating sessions with storyboards}
For the one-on-one storyboard speed dating sessions, which lasted about 30 minutes, we reminded participants of the scenario from the focus group sessions (to imagine they were recovering from medical treatment or surgery). Participants were shown each storyboard and then interviewed about their emotions and reactions. 

\subsubsection{Demographic questionnaire and compensation}
Each participant completed a demographic questionnaire online after completing all study sessions. Participants were then sent \$25 gift cards per session they participated in as compensation.

\subsection{Analytical approach}
All data was transcribed from video recordings using Zoom's built-in transcription feature and then analyzed using thematic analysis as described by \cite{braun2006using}. During \textbf{phase one} of analysis, we familiarized ourselves with the data by reviewing the transcriptions and referring back to the videos when the transcriptions failed for an overview of the main discussion points of each participant and group. These broad concepts served as the foundation for the first set of codes used for interpreting the text during phase two of the analysis. These codes included: adaptable, feedback modality, Black culture, human agency, technology is fine, preference for humans, motivation (content), motivation (format), age, disability, and privacy concerns. 

During \textbf{phase two}, we annotated transcripts using a combination of the initial list and open coding, updating the code list to reflect new findings as they arose. For example, we added a code for technology acceptance and one for ability hierarchy. For \textbf{phase three}, the codes were categorized into main themes and organized with supportive text. The themes included design of the curriculum's content, data collection modality, accountability concerns, future use of the data collected, Black community standards, Black culture, choice of language, and inclusivity. The themes were developed through an iterative process of categorizing codes that had similar tones, were nuanced, and provided meaningful insight relative to the research questions. 

\textbf{Phase four} involved completing a detailed written analysis of the themes and the supportive text, which was iteratively refined as the ideas and narrative came together. The transcripts were double-coded by two Black researchers as they are familiar with African-American English (AAE), which was frequently used in the interviews.

\subsection{Researcher positionality}
All interviews were conducted by the second author, a Black researcher. The data was reviewed by both the first author, who is also Black, and the second author. This method was used to recognize cultural competencies and experiences, providing a deeper understanding of shared experiences beyond words alone \cite{ogbonnaya2020critical}, 
and responds to recommendations about how to better interpret responses by understanding the perspectives, experiences, and backgrounds of the population \cite{harrington2019deconstructing}. Three additional races/ethnicities are represented in our research team. This work is a result of rich discussions, informed by our lived experiences and diverse professional expertise (e.g., medicine, engineering, human-computer interaction, and design), about how to conduct the study, analyze the data, and interpret the findings for a wide range of potential future uses. 

\section{Findings}
We describe our findings by the research questions, covering participants' motivation to use the \textit{Care Curriculum}, their perceptions of the technology and its opportunities and challenges. 

\subsection{RQ1: Motivation} 
We outline findings regarding the \textit{Care Curriculum} content from a healthcare angle, the amplified need for the \textit{Care Curriculum} expressed by those who could empathize with not having a caregiver, and the requirement for the \textit{Care Curriculum} to be adapted to an individual's needs.

\subsubsection{Care Curriculum content}
Participants noted that Nana received minimal information about what to expect from the curriculum. 
Most felt that having an orientation at the beginning of the program would be beneficial. Some mentioned that the orientation needed to be highly informative and allow older adults to ask many questions if needed. Most participants responded positively to the curriculum's content that aligned with the general idea of recovery, such as goal setting, fall prevention, nutrition, and physical activity. The \textit{Staying Safe} storyboard introduced a training video to inform about avoiding falls. Nana is depicted as sitting and listening to all of the information in one session and feeling informed after completing the training. P4 stated that he was more inclined to follow a training video on falls, as a care provider would provide less factual information in this instance. Additionally, P7 expressed that she would listen to the training video for information, citing her growing concern about falling as she gets older. However, P3 felt elements of this storyboard were unrealistic, as he believed the older adult would need time to complete the training throughout the day, stating, \textit{``This struck me as being outside of the normal. If [Nana] had said get back to me in about half an hour and I will [complete the training], then that would have been realistic. I'm surprised that she completed the training.''}

There were some differences in opinions about the effectiveness of certain storyboards in supporting recovery. For example, the \textit{Feeling Fatigued} storyboard shows Carol suggesting that Nana dance when she notices Nana has been more fatigued than usual. Many participants felt it was positive that Carol pushed Nana to move although she was tired; however, P9 stated, \textit{``I don't know how realistic [this storyboard is for] someone [who] is really hurting or fatigued.''} Additionally, P23 stated it was important that Carol gave Nana an option to either dance or continue to rest if she was extremely fatigued. Another difference in opinions came up in the \textit{Positive Reminiscing} storyboard, which is a moment between Carol and Nana where Carol suggests that Nana recall positive memories. Some participants felt this element may not make sense as content for the \textit{Care Curriculum}, as it could not positively impact their recovery. P23 stated, \textit{``Although it seems like it's positive, it may [make them feel down],''} or should be left as a moment between family and friends, as stated by P17, \textit{``Yeah, when you reminisce, you wanna talk to a personal friend or someone close to you that can hear you, or even respond to your memories.''} Similarly, P14 stated, \textit{``I would rather reminisce with a human.''} However, others felt it was great, P26 stated, \textit{``This [storyboard], to me, addressed the emotional [well-being aspect of recovery]}.'' Additionally, P16 expressed that positive reminiscing helps to alleviate stress and P7 mentioned that since Carol is not a human, she would be willing to actively listen for as long as Nana needed. The findings indicate that there are certain fundamental aspects of recovery that should be included in the curriculum as baseline components (e.g., orientation, physical activity, and goal setting), while other elements (e.g., training, persistent motivation to stay active, and positive reminiscing) should be tailored to the specific needs and preferences of the individual patient.

\subsubsection{Personal exposure to a lack of caregiver}
Most participants showed a greater willingness to engage with the \textit{Care Curriculum} when it was evident that the program could assist individuals known to them who could benefit significantly, usually those with high degrees of frailty and without always-available caregivers. For example, during the speed dating sessions, certain storyboards were seen as positive as many participants stated that aspects of the curriculum could be beneficial in their previous post-hospitalization and recovery experiences, in supporting older relatives, and in assisting residents of assisted living facilities where they indicated they had experience. P14, stated: \textit{``I had a knee replacement surgery [...] right when Covid hit and [physical therapy] shut down, so I had to do [physical therapy] at home. So something like this would have been very helpful for me [...] to have had after that surgery.''} P14 also anticipated that the \textit{Care Curriculum} could support her daughter in taking care of her in the future. P23 expressed, \textit{``I would use this program [based on my experience] in caring for my mom.''} P23 acknowledged the program's relevance based on his caregiving experience for his mother, indicating his intention to use the IVA program personally in the future if ever in a similar state. P23 also stated, \textit{``I’m also concerned about other seniors and people that I meet who do not have a caretaker there at home.''} P4 shared, \textit{``So, my mother is navigating through all of this, and she has an Alexa device. [Therefore, I would use it.]''} P4, having a mother in need of care and already familiar with Alexa, saw the potential of this system to fill some of her care needs. 

\subsubsection{Needing an adaptable Care Curriculum}
All participants expressed a preference for a system that was inclusive of people of all limited abilities. Every participant positively perceived the \textit{Physical Activity} storyboard. In this scenario, Carol suggests an activity that Nana can do given her change in ability, due to pain, and walks her through a 10-minute bed exercise that is tailored to her pain. All participants agreed that proposing modifications so that Nana could continue with her exercise was highly acceptable. P16 stated \textit{``This [system] is a very helpful tool for when you're limited''}. Additionally, many participants expressed a desire for the system to consider their daily needs and provide the flexibility to choose the level of assistance they require. For example, P12 commented, \textit{``It’s nice that adjustments can be made to [cater to] how Nana may feel on a given day.''} P12 also elaborated on the types of adjustments she would want, saying, \textit{"I might be lazy today and need some initiative to help get past that, or my back is really [hurting] today and I want to do something, but not what I normally would do."} P3 stated \textit{I think if technology is not flexible and becomes too rigid for the user, it decreases its usefulness.''} P3 links interface adaptability to usefulness, highlighting the need for the curriculum to be tailored to different levels of care. P3 also stated, \textit{``Anything that helps you solve a problem is an ally. I look for tech to help solve a problem. The system made [Nana] feel better which increased the buy-in.''} P1 mentioned his lack of confidence in systems that have not successfully adapted to his needs. This feedback suggests that successful system adaptability is crucial for user engagement and trust. 

While all participants expressed support for an inclusive curriculum, many were skeptical of their likelihood of using it. This reluctance stemmed from their belief that the curriculum was unnecessary for people with high activity and ability levels, such as themselves. They indicated that the curriculum was an effective option for those who \textit{``needed it,''} distancing the system from themselves. For example, during a focus group session, P10 stated \textit{`` I'm not sure [the Care Curriculum] should just be for older people, there are many disabled children, young adults, older adults, who might be able to benefit from something like this, \textbf{but for able body people at our age, I'm not a fan.''}} In this comment, P10 is suggesting that she sees people with disabilities as the target group of the \textit{Care Curriculum}. P9 added, \textit{I could see how [the Care Curriculum] might be useful for people who are not at our level of functioning''} to the discussion. 

A similar idea was present in the storyboards. In the \textit{Healthy Weight} storyboard, Carol notices that Nana has gained weight, calls her doctor, and schedules an appointment. Many participants were unable to visualize themselves at a level where they could not schedule their own doctor's appointments. P8 stated, \textit{``It’s too bad that Nana can’t make her own appointments, but obviously she’s at a level where she needs that kind of assistance.''} P15 believed that Nana should have been aware of her weight gain and taken steps to address it before Carol intervened. This statement suggests that P15 found it difficult to imagine herself in a situation where she wouldn't notice her own weight gain. P9, again, also distanced herself from those who need assistance. Despite having had experience with knee surgery recovery, P9 stated, \textit{``I would probably not be in that position, because I move all the time.''} Many participants perceived that the curriculum having advanced capabilities and features (e.g., a camera and doctor communication) is only necessary for people of extreme need and introduces a stigma related to ability and aging. For example, P8 stated,\textit{``If I was in Nana’s position and needed that level of support, I’d be fine with Carol being able to see me visually.''} Additionally, P17 stated \textit{``It's a good idea for those who are not able to,''} referring to the ability that Carol has to notice Nana's weight gain and call her doctor. 

Most participants had positive feedback and expressed a need for features of the curriculum that aligned with their current ability. For example, P1 noted that he liked that the \textit{Care Curriculum} included reminders for Nana to complete tasks and expressed that current systems that he uses does not have this feature. P8, P12, P16, P25, and others also stated that having a curriculum that had reminders would encourage them to use it as they currently need them. Several participants expressed a desire for a system that could provide additional motivation and encouragement to complete tasks at a higher level even when they are struggling or in a difficult state. The goal would be for this system to help them reach a point where they no longer require the structured curriculum and the assistance of Carol.

\subsection{RQ2: Perception of the technology}
We now present our findings on three key areas: 1) consentful interactions, 2) IVA companionship, and 3) enhancing engagement and health management through IVA.

\subsubsection{Consentful interactions}\label{sec:consentfulinteractions} 
For some scenarios, the storyboards showed Carol engaging Nana in the curriculum without first explicitly asking for Nana's permission. Other scenarios depicted Carol ``noticing'' or ``seeing'' Nana's condition, and then proposing a customized curriculum to specifically support Nana's needs. Considering these scenarios, all participants expressed a preference for a system in which Carol sought consent from Nana before initiating any program, with many participants expressing concern about Carol ``noticing'' or ``seeing'' them. P3 stated, \textit{``I think it was important that she ask Nana. […] I’m assuming she would continue to get Nana’s buy-in at each step.''} For P3, consent was not only needed but needed on an ongoing basis throughout all interaction steps. P7 similarly stated the need to be asked for consent but did not specify how often, \textit{``I think maybe [Carol] needs to ask me first.''} P4 echoed the need to give permission in reference to the \textit{Physical Activity} storyboard, \textit{``I’d love that kind of feedback. I [would] probably utilize that feedback to motivate me, […] as long as I’ve initiated the conversation or the session.''} In this case, initiating the session is a form of granting the IVA consent. 

The concept videos depicted scenarios where the older adult, rather than the IVA, would initiate the curriculum-based interactions. In these depictions, the participants did not express a need to obtain consent before engaging in the activities. This observation suggests that consent was implicitly granted by initiating the interaction. As a whole, we found that all participants wanted to give pre-approval before Carol engaged with them, highlighting their desire for autonomy and control over the interaction, whether it was an ongoing process, before interacting with the IVA, or implicit through user-initiation. 

\subsubsection{IVA companionship}
All participants held diverse perspectives on their perceptions of IVAs in this context. Some participants viewed Carol as a potential companion, valuing the positive aspect of having someone to talk to, as expressed by P25, \textit{``it’s always positive when you have someone to talk to.''} Additionally, a few participants saw Carol as a potential substitute for those who lacked a support system. \textit{``If I didn't have any family or have anyone checking on me, I would appreciate some intervention,''} P9 stated. While some other participants were hesitant about using Carol for companionship, P17, for instance, expressed skepticism, stating, \textit{``I don't know if I would go into details talking to a computer because the computer really doesn't have the emotional [capability] to [respond].''} A few participants openly expressed that they did not view Carol as a companion, for example, P15 stated \textit{``I don't feel friendship with [Carol].} 

However, some participants were willing to reconsider their initial reservations about Carol as a companion, influenced by the positive outcome observed in the storyboard scenarios. P26 shared, \textit{``I don't see [Carol] overstepping [in] this particular case. [Nana] feels more grounded and more in control of her emotions, so apparently, [she] benefited from the suggestion. I think that’s good. I think sometimes we need somebody to kind of pull our chain and say, you know, you need to kinda chill out a little bit.''} These findings highlight the complexities of participants' attitudes toward a voice assistant's potential companionship role, emphasizing the importance of a nuanced understanding of user preferences and experiences in the development and implementation of these technologies.

\subsubsection{Enhancing engagement and health management through the IVA}
Some participants believed that the IVA's tone of voice should match the content presented in the curriculum. P16 expressed, \textit{``If the voice was conducive to what you're doing [I would use this].''} P19 suggested the IVA have a calming voice for activities like breathing exercises and P16 recommended a lively voice for more energetic activities. This finding indicates that tailoring the voice to the content can enhance user engagement and create a more immersive experience. Most participants felt positive about integrating IVAs into recovery care. For example, P26 stated, \textit{``This is a great way to reinforce and to help guide the person through what they're supposed to do after they leave the doctor.''} P3 expressed a similar thought, \textit{``This is a case where, having an inanimate object may be superior to having a human being,''} suggesting that an IVA is the better modality for delivering recovery care. 

Several participants noted that the IVA's functionality could extend beyond the scenarios of the \textit{Care Curriculum}. P24 and P26 stated that they see an IVA as being beneficial for relaying doctor's summaries and written instructions after visits, as well as providing assistance with managing prescription medications. Additionally, P5 expressed a desire for an IVA to communicate health trends and information about their condition, which could aid in making informed decisions about seeking medical attention. These perspectives highlight the potential to incorporate overall health management, decision-making, and adherence to treatment plans in IVA-based recovery care.

\subsection{RQ3: Opportunities}
Many participants voiced aspects to consider and opportunities to improve the \textit{Care Curriculum}. This section describes our findings about 1) racial identity and healthcare, 2) enhancing the \textit{Care Curriculum} with foundation models and artificial intelligence (AI), and 3) IVA pervasiveness within the home.

\subsubsection{Racial identity and the Care Curriculum}
Several participants expressed that Black older adults adhere to a similar set of principles, guided by their racial identity. For example, P6 said, \textit{``We express ourselves differently than other ethnic groups,''} referencing the differences between Black people and other groups, namely white people. P26 noted that \textit{``Our physiology is different than other races, and where they think cholesterol is high for white folks, it is a normal baseline for us,''} highlighting the connection between Black people in comparison to other racial groups. Note, there is scientific evidence showing that cholesterol affects Black and white people differently, creating certain risks for white people that are not as prevalent for Black people and creating a need to distinguish cholesterol types further \cite{tejera2021high, woudberg2016protection}. This finding suggests that the \textit{Care Curriculum} should be scientifically and medically accurate by considering nuanced differences in health regarding the racial identity and background of the user. 

Most participants further identified specific moments where the example scenario seemed more geared toward a white audience. For example, regarding the \textit{Physical Activity Coach} video scenario, P5 stated, \textit{``I don't think I would like to exercise to Pride and Prejudice or The Great Gatsby [...] and most Afro-Americans, I would suggest, may feel the same way I do about those.''} P1 also stated, \textit{``Yeah, I noticed none of the books were Black books.''} In this quote, the term \textit{``Black books''} is used to refer to literature that is either written by, about, and/or widely accepted by Black individuals. The other participants in this focus group agreed that a representative sample of books would be most appropriate, not just books that are deemed \textit{``classics.''} Most participants perceived that a more culturally responsive curriculum that reflected the preferences of a Black audience would be more effective, either through the communication style of the curriculum, the measurement of health, or the motivational content.
 
\subsubsection{Enhancing the Care Curriculum with foundation models}
Most participants showed interest in the capabilities of foundation models and responded positively to potential machine-learning interventions in the \textit{Care Curriculum}. P3 stated, \textit{``[...] it shows that the technology is storing information from past interactions and using that to take further steps to enhance the therapies that they've already offered to Nana. I think this is maybe another good thing with technology.''} Similarly, P25 expressed, \textit{``I'm having a conversation with the machine [and] once I tell her about my various experiences, I'm sure she would have some questions to further categorize.''} P23 expressed regarding Carol detecting Nana's stress in the \textit{Healthy Mind} storyboard, \textit{``If the system can detect that, I think that's a good thing.''} Regarding potential applicability, P26 stated, \textit{``I see value in this. As you get older the print gets smaller [...]. I could hold it up to the machine or type in [that I am] taking this medication and it [gives] you the direction from your doctor.''} Many participants used their knowledge of AI, algorithms, and machine-learning to suggest approaches that could influence the effectiveness of the \textit{Care Curriculum}, reflecting a favorable view of the incorporation of such systems. For example, P25 expressed, \textit{``The algorithm is written so that [it] can ask more questions, or make a more effective bank of knowledge to help me later on.''} Additionally, P26 stated, \textit{`` I'm sure this is all new AI technology or expanding AI technology; so does this system allow the person to choose what their reward system is like?''} This comment suggests that incorporating AI would allow more personal choices in the curriculum. 

Many participants expressed a level of comfort and trust with AI use in IVAs. P25 stated, \textit{``I mean, yeah, it's a machine, but that's where technology is going. Algorithms are getting so complicated that you sit down and have a conversation with a machine and not feel that you're missing anything.''} P1 expressed, \textit{``it's interesting [that they] find out how we work.''} P9 stated, \textit{``Carol is very perceptive. Carol is actually like a computer or AI. [...] so I think [it's positive] that Nana has been willing to accept all of this advice from her up until this point and that she is grateful and appreciates [her help].''} P4 expressed, \textit{``[...] instead of paying for a therapist, I can talk to a machine. I [would] probably do that.''} These findings suggest that many participants understood and embraced the idea that technology, and AI in particular, is an integral part of life and presents an opportunity to address gaps within the \textit{Care Curriculum}.

\subsubsection{System pervasiveness within the home}
As all participants asked questions on Carol's machine learning capabilities, several also alluded to the idea that Carol may need to move around or be more pervasive in the physical space to provide continuous support. P4 noted, \textit{``The only thing about this [system] is I have [to have] the device move from whichever room it was [in] before, [...] which is likely a kitchen or a dining room to my bedroom. So either we have multiple devices or I gotta move the device.''} Additionally, P23 questioned Carol's ability to obtain information in various rooms \textit{``[...] if [Carol is] upstairs in the bedroom, and the box is up there, [then] how does [Carol] know when [Nana is] going down the stairs, and therefore having a balance issue.''} These findings suggest that the system's placement, feasibility of relocating the curriculum within the home, and integration to other sensors within the home and communicating this integration are crucial factors to consider when designing the \textit{Care Curriculum}. 

\subsection{RQ3: Challenges}
Now we present our main findings on the challenges of creating an inclusive \textit{Care Curriculum} that effectively addresses diverse user needs and preferences. We describe findings related to the challenges associated with 1) marginalization by race and socioeconomic status, 2) representing everyone, and 3) accountability and privacy.

\subsubsection{Marginalization by race and socioeconomic status}\label{sec:marginalizationbyraceandsoes} 
Most participants recognized the significance of inclusivity for individuals with historically marginalized identities. P6 expressed a perspective shared by other participants stating, \textit{``The bulk of Black Americans live in urban areas and then rural areas. How [internet] connection is done, in terms of connectivity, Black people live where we're kind of shut out of technology.''} Black people that live in certain areas are excluded from technology due to limited internet access. This phenomenon is known as digital redlining, classified as the intentional exclusion of certain groups from internet access \cite{mccall2022socio}. Research has shown that digital redlining predominately affects marginalized communities like Black individuals, older adults, and people who live in rural areas \cite{mccall2022socio}. 

A more nuanced perspective of inclusion emerged as many participants further discussed various forms of marginalization inherent in the \textit{Care Curriculum} and its format. For example, there was an acknowledgment of the need for technology that was designed to be inclusive of all members of the Black community by considering individuals from all socioeconomic levels. This recognition stemmed from the perception that the design would not appeal to people with lower education and income levels. When discussing the selection of audiobooks, P9 emphasized, \textit{``they're not easy reads,''} emphasizing the need for content that is more accessible to a broader audience with lower education levels. 

P26 brought attention to the importance of including people from diverse geographic regions in the presentation of certain aspects of the curriculum, stating, \textit{``it depends on the audience you’re trying to reach,''} with a specific reference to those residing in the Southern United States. P6 further specified communities that often face exclusion from programs like the one under consideration, stating, \textit{``but people like me who grew up in the hood [would not be able to access this].''} In this context, P6 uses the term ``the hood'' to characterize neighborhoods where traditionally marginalized and low socioeconomic status communities reside. Some participants raised questions about cost and its importance to older adults. P1 stated, \textit{``[For] most seniors, the first thing they [are] gonna ask is how much does Carol cost? Is Carol gonna be covered by my insurance?''} The participants' perspectives highlight the complex nature of inclusivity in healthcare technology, emphasizing the importance of addressing not only racial and age-related disparities but also social barriers that can prevent access to and engagement with digital health solutions such as the \textit{Care Curriculum}.

\subsubsection{Challenges of representing everyone}
Most participants expressed a preference for Black cultural representation in the IVA, as P7 stated, \textit{``I think I would just like to have someone that’s more in tune. I know you can’t have a Black person talking to every group on this voice assistant, but [I want] someone who can be more alert, more in tune, [and] more targeted.''} P14 more specifically said, \textit{``I don't think I want to hear a white person in my house all day telling me what to do,''} referring to the IVA's traditionally-white accent. P15, a woman, shared \textit{``I would like to see [...] some Black female figures exercising their voices,''} underscoring the desire for representation as part of the IVA. 
Simultaneously, we noticed behaviors and perceptions that may create challenges for inclusive IVA design. For example, while some participants did not want the IVA to sound white, they were also not receptive to the actresses' Latinx ethnicity. P20 expressed, \textit{``She's speaking English, but not clearly enough''} regarding what P4 labeled as \textit{``a little bit of a Spanish accent.''} This sentiment was echoed by many participants. P25 said, \textit{``In the South, we are not going to understand Hispanic dialects.''} Additionally, P7 said \textit{``She throws me off. I think it's the fact that she has [...] a Latin tongue.''} However, during this discussion, P9 visibly appeared uncomfortable with the criticism other participants were directing at the voice used in the system. In an effort to redirect the conversation, P9 encouraged P7 to focus on the system itself rather than the Latinx voice, saying, \textit{``If you try to remove that person from it, and just concentrate on Alexa, I can see a need.''} This intervention by P9 suggests that the opinions expressed about the voice were not universally shared among all participants. It highlights the diversity of perspectives within the group and indicates that some participants may have seen value in the system beyond the specific voice used.

A few participants shared a perspective about acceptable options that reflect perceptions of Blackness. P5 indicated a preference to replace the woman in the video with a person like Redd Foxx, who was a well-known Black American comedian and actor. P5 explicitly mentioned, \textit{``I would replace her with Redd Foxx.''} P5 desired a character who was relatable to him (regarding racial and cultural background) and had a personality similar to Redd Foxx's stage roles. For example, he expected the character to bring humor by engaging in physical activities that highlighted a limited exercise ability. Including a character with a humorous nature was viewed as a motivator, with him expressing the sentiment that witnessing the character's actions would inspire him, thinking, \textit{``if he can do it, I can do it.''} In contrast, in the same focus group, P6 opposed the inclusion of Redd Foxx, stating, \textit{``I’m not going to go as far as Redd Foxx, but I agree.''} He aligned with the idea of replacing the woman in the video with a more relatable person for demonstrating activities but did not consider Redd Foxx to be a good choice, given the actor's extreme personality. This dialogue emphasizes a gradient in the perception of acceptable representation of Blackness and cultural inclusion in the curriculum. 

\subsubsection{Perceptions of gender and social norms} Some male participants spoke derisively of the idea of sending one another flowers. P3, a man, expressed, \textit{``I didn’t appreciate the garden analogy. It would have been better to get something I could relate to ... maybe a ball game [or] a home run.''} P21 spoke similarly in another focus group, \textit{``I play basketball. So for me, [...] as I shoot free throws and make free throws, [I feel] I’ve accomplished something instead of a garden.''} These statements indicate that some participants aligned their gender identity with aspects of the curriculum, seeing flowers as not masculine and expressing a preference for features they associated with masculinity. These associations extended to professions. P4 stated, \textit{``for ... an engineer, you wanna report that you did your exercise... as opposed to [sending] somebody a flower.''} While this quote betrays a narrow view of engineers, it also demonstrates the wide range of preferences the curriculum must cater to. However, there were women who also did not align with the garden analogy. P9 expressed her dislike for the garden analogy and P20 related the garden to aspects of a casino theme, aligned with games that she preferred to play. These findings indicate that the complexities of preference and that all analogies may not universally resonate with all individuals.

\subsubsection{Accountability and privacy}
Many participants voiced concerns regarding the accountability for medical injuries resulting from IVA recommendations. P1 noted, \textit{``What liability does Carol have? Suppose Carol suggested I do something, and I did it, and I seriously injured myself.''} P16 specifically was concerned about a potential scenario in which she got hurt while alone and following Carol's advice \textit{``'cause there's nobody there with me,''} to emphasize the problem surrounding the perceived lack of accountability. Several participants also expressed concern about where and how their data would be used. For example, P12 shared that the IVA might \textit{``gather information about me and sell it or share it with someone that I wouldn't want to.''} P12 also stated, \textit{``The only concern that just popped in my mind for this would be if there's some bias in how the information is taken by Carol that is cultural.''} P15 questioned \textit{``Where will this data be going after?''} These findings must be considered while acknowledging past unethical research practices that have particularly impacted Black communities, including the incident of Fannie Lou Hamer who received a non-consensual hysterectomy \cite{volscho2010sterilization} and the Tuskegee Syphilis Study when over 300 Black men were intentionally not given penicillin to treat their syphilis \cite{brandt1978racism}. These statements carry the weight of this history for our Black participants.

\section{Discussion}
In this study, we explored the perception of Black older adults regarding the use and factors that would motivate the use of the \textit{Care Curriculum}, IVAs enacting the curriculum, and the challenges and opportunities they anticipate. We now discuss our findings through three main angles: 1) navigating inclusion in the healthcare framework, 2) the \textit{Care Curriculum} in the current healthcare systems, and 3) the future of the \textit{Care Curriculum}. Additionally, we provide a list of design recommendations, as shown in Table \ref{tab:recs}.

\subsection{Navigating inclusion in the healthcare framework}

We now discuss the challenges of inclusion regarding the \textit{Care Curriculum} with particular attention to inequities, disparities, infrastructure gaps, and harms we have inherited from the past. 

\subsubsection{Addressing inequities}
This research is significant because it is one of the few studies focusing on an interactive, automated medical recovery curriculum tailored specifically for this demographic. According to our participants, the need for and use of the \textit{Care Curriculum} was not solely based on physical and cognitive capabilities for Black individuals, but also determined by inequities in healthcare more broadly, such as limited and unfair inclusion of Black people in medical research. The findings show that these inequities are caused by factors such as health disparities, socioeconomic status, geographic location, historical exclusion from technology, healthcare, and systemic racism. The findings also suggest that the curriculum can address healthcare disparities by providing a path to access to personalized care, care recommendations aligned with the user's background and needs, and health resources while empowering people to make more informed healthcare decisions. Additionally, the curriculum has the potential to serve as a resource for advocacy, particularly for marginalized groups, by providing evidence-based data to support their health concerns. The \textit{Care Curriculum} should be adaptable to the diverse needs and circumstances of older adults, considering not just their physical health but also the broader social and systemic factors that shape their experiences and access to care. 

For our participants, the quality of the \textit{Care Curriculum} is intrinsically linked to the fairness and representation of their cultural and racial identity, which is crucial for building trust and motivating continued use of the curriculum. The participants expressed a need for the curriculum to address them. Therefore, by first acknowledging the deep-rooted connection between healthcare, culture, and race, and then by designing recovery programs that are culturally responsive and representative, healthcare providers can foster greater trust and engagement within Black communities. The \textit{Care Curriculum} should serve as a blueprint for an ideal healthcare system, rather than just replicating the current one. Curriculum designers should adopt a continuous and iterative approach, focusing on envisioning and implementing improvements that address existing shortcomings while incorporating innovative practices, ultimately leading to better patient care and outcomes. Our research contributes to the literature by emphasizing the importance of incorporating racial health disparity factors in the \textit{Care Curriculum}, recognizing them as genuine healthcare concerns. 

\subsubsection{Navigating infrastructure gaps}
It is also important to consider how socioeconomic status and geography may impact access to the \textit{Care Curriculum}. Although data sources indicate that older adult ownership and use of mobile devices (e.g., smartphones and tablets) and ``smart TVs'' is steadily increasing \cite{RN501}, the cost may place them out of reach for low-income users, especially among families coping with medical costs associated with serious illness. No or low-speed internet access may be a concern in rural and low-income areas. As stated by our participants, cost may be a concern for older adults, while internet and technological access may pose a barrier for Black older adults (see Section \ref{sec:marginalizationbyraceandsoes}). Research has shown that half of US counties do not meet the federal threshold for up- and download internet speeds and are affected by poor urban infrastructure due to ``digital red-lining'' \cite{RN502, RN503}. Our participants pointed out these gaps, some from their own lived experiences growing up in ``the hood.'' As Internet availability affects access to high-quality healthcare interventions through telemedicine and IVA use cases such as the \textit{Care Curriculum}, connecting research and design-related evidence with policy initiatives may drive change. Federal and state initiatives to extend broadband infrastructure and provide subsidized or free internet services, low-cost devices, and training have great potential to improve educational and population health outcomes, but are not yet universally embraced \cite{bauerly2019broadband, RN502}. Developing ready-to-use adaptable programs like the \textit{Care Curriculum} may increase the impact of these programs, and therefore also increase the incentive to embrace them.

\subsubsection{Acknowledging past harm in digital interface privacy practices}
It is critical to acknowledge the historical harm experienced by the Black community \cite{ray2022going} in the realm of data sharing and collection. To address this, there should be transparency regarding the information collected, how it is stored, and the intentions behind collecting such data. This transparency empowers users to make informed decisions about their comfort level in using the \textit{Care Curriculum}, and to advocate for change when invasive data collection may hinder them from benefiting from it. The liability of the IVA should also be transparently communicated. If patients understand that the IVA cannot be held accountable for the recommendations it provides, they may be more inclined to accurately communicate their pain and ability levels, ensuring that the curriculum is tailored to their specific needs. By fostering transparency in both cultural representation and data practices, the \textit{Care Curriculum} can be designed and used in a manner that respects and aligns with the values and concerns of Black older adults. 

\subsection{\emph{Care Curriculum} in the current healthcare systems}
In this section, we discuss how the \textit{Care Curriculum} fits into current healthcare systems. We discuss our findings related to the intersectionality of age and ability, the continuous availability of the system, and personalization. 

\subsubsection{The intersectionality of age and ability.}
Most participants strongly emphasized the need to distinguish between aging and ability, advocating for the \textit{Care Curriculum} to accommodate this distinction. Previous studies have similarly highlighted the importance of recognizing that age does not necessarily correlate with one's ability, especially when implementing innovative strategies for older adults \cite{knowles2021harm, kim2021exploringOlder,  o2020voice}. \citet{knowles2021harm} argues that age and ability are distinct concepts, and conflating the two can lead to the assumption that disability is inevitable in old age. This misconception may result in older adults without disabilities or those with certain high-functioning disabilities being excluded from access to products designed for their specific needs. Consistent with this idea, our study participants showed decreased motivation to accept or implement the \textit{Care Curriculum} if it focused on reducing frailty, as they found it difficult to envision themselves needing such assistance. Participants indicated that if they found themselves in extreme need, meaning they had lost significant ability to support themselves independently, they would be more open to using the curriculum and be accepting of more invasive methods of delivering it. However, some of the participants' relatively high socioeconomic status may have contributed to their difficulty visualizing themselves as needing an automated curriculum, as they currently benefit from their social status and could potentially have access to paid caregivers in times of need.

One design possibility involves harnessing participants' optimism toward leveraging LLMs and AI in the \textit{Care Curriculum}. Advanced technology may engage Black older adults in innovative ways that overcome the stigma of aging and ability, as well as changes in ability based on social factors and background. Participants indicated curiosity and a desire to be part of the technological advancements using LLMs and AI. Therefore, increasing usage is possible by creating advanced products that cater to this population. However, incorporating these advancements should be done carefully as their inclusion can also potentially negatively impact the healthcare of Black populations. A 2023 study by \citet{omiye2023large} found that LLMs can perpetuate racist ideas about the pain tolerance of Black and white individuals. Some models were observed to propagate the idea that Black individuals feel less pain or are less likely to report pain compared to white individuals due to being perceived as “weak” \cite{omiye2023large}. As a result, \citet{omiye2023large} advises caution when integrating LLMs into healthcare systems, urging consideration of these harmful beliefs. Based on the findings, adopting an inclusive approach that considers users' diverse experiences may motivate historically marginalized individuals to engage with and benefit from these programs. However, the findings also raise questions regarding the sensing and data capturing necessary to make the \textit{Care Curriculum} adaptable and what adaptations are ethical/unethical. 

\subsubsection{The continuous availability of the system}
Our participants highlighted the importance of the \textit{Care Curriculum} for older adults who lack a caregiver. Access to licensed in-home caregivers can be prohibitively expensive \cite{stone2021developing}. Additionally, caregiver availability can be strained due to healthcare demands, leading to gaps in care \cite{jones2024healthcare}. Even those with family members who serve as their primary caregivers experience interruptions in care as family caregivers balance various tasks and roles \cite{hazzan2022family}. The \textit{Care Curriculum} can address these issues in several ways. For older adults requiring recovery care, it can serve as an always-available source of care, complement part-time licensed caregivers to enhance efficiency, or act as a tool for family caregivers seeking guidance on healthcare-related decision-making. The system allows caregivers to interact with the IVAs and \textit{Care Curriculum}, keeping them updated on the older adult's recovery progress, highlighting areas of success or struggle, and offering suggestions for improvement. Participants expressed interest in advanced technology and machine learning algorithms to provide and display information about their health and medications. These algorithms could also offer caregivers guidance on effective care strategies for older adults in recovery. This comprehensive system aims to bridge care gaps and enhance the overall quality of care for older adults, whether they have full-time, part-time, or no caregivers.

Regarding expanding the \textit{Care Curriculum} to other areas of health management, the system could potentially serve as a virtual caregiver through its content and the IVA. For example, the system could include advanced features beyond recovery care, such as medication reminders. These reminders could be implemented with increasing urgency: starting with subtle sounds or dings 15 minutes prior to medication time, and progressing to spoken reminders if the older adult has not acknowledged taking their medication after some time. The system could also use the IVA to detect social isolation by identifying limited voices in the home. In response, with the use of advanced algorithms, the curriculum could provide increased elements of well-being, such as encouraging more modules of positive reminiscing and suggesting less rigorous physical activities to ensure safety for those spending most of their time alone. These features would allow the \textit{Care Curriculum} to become an integral part of the older adult's daily life during recovery, potentially filling critical gaps in care. 

\paragraph{A community orientation towards helping one another}
Our participants also shared the system’s relevance in supporting their family members who either needed care or support in caring for them. This finding highlights the opportunity to incorporate connected features that involve the family and friends of the user. Integrating elements that engage the broader support system can increase motivation and sustained engagement with the program among Black older adults. This finding underscores the value of designing healthcare interventions that recognize the importance of social connections in enhancing the user's experience and commitment to the program. This finding is consistent with previous research indicating that incorporating community connections in IVAs \cite{kim2021exploringHow} and care practices \cite{harrington2018informing, chaudhry2016successful} can enhance engagement. Considering this community orientation, it will be key to take older adult's social support network 
\cite{wang2024understanding, chaudhry2022formative} into account in the design of the system, such as by allowing for layered access to different people. For example, trusted friends could be notified when a select number of activities, such as a stretch or balance exercise, are completed, so that they can provide encouragement to the older adult. 

\subsubsection{Personalization}
The \textit{Care Curriculum} should aim for inclusivity, catering not only to Black older adults but to a diverse range of users across all demographic spectrums. When offering digital content such as books, podcasts, movies, or music, it is crucial to recognize that shared characteristics like race, age, or gender do not guarantee uniform preferences. While individuals may share certain experiences based on these factors, it is important to avoid generalizations and instead provide personalized content that respects the unique interests of each user. This approach acknowledges that Black older adults, like any group, are not a monolithic entity but rather a collection of individuals with varied preferences and experiences. This personalization might be possible by linking to existing systems with a vast number of options that have become extremely capable of delivering personalized recommendations, such as YouTube, Netflix, or Spotify. 

Similarly if using metaphors, the design must consider the social signifiers those metaphors contain for different people. To some of our participants, the garden metaphor did not resonate the way a metaphor associated with sports such as baseball or basketball would. A possible technological solution is to customize and tailor these sorts of metaphors to specific individuals by using AI to dynamically rebuild systems based on them. Customizing the \textit{Care Curriculum} to diverse preferences presents an opportunity to increase usage across a broader population. 

The curriculum's content should also be personalized to align with the specific type of recovery and the aspects that are most relevant and meaningful to the individual patient. For example, patients with strong support systems can reminisce with family and friends, so their emotional well-being content might differ from those without such support. Patients without strong support may prefer to engage in reminiscing activities within the curriculum itself for emotional well-being. However, for patients experiencing severe depression-like symptoms, certain aspects of the curriculum could potentially trigger or exacerbate depressive episodes. Therefore, it is necessary to develop a comprehensive understanding of the participant's patient profile while tailoring the curriculum. The patient should be actively involved in the decision-making process and have the opportunity to provide input regarding the various content elements. 

Moreover, the personalization aspect should not be a one-time function at the beginning of the process. Instead, the system should have the capability to continuously personalize and adapt the curriculum throughout the recovery journey. This dynamic personalization should be based on ongoing information obtained directly from the patient, as well as insights gathered from follow-up physician visits. During these follow-up appointments, discussions should include an evaluation of the curriculum and recommendations for further personalization to optimize care and facilitate the patient's recovery.

However, given that the \textit{Care Curriculum} is intended to support recovery, certain aspects should be consistent for all patients to prevent exacerbating disparities in recovery outcomes. Therefore, guidelines should be developed to ensure the \textit{Care Curriculum} maintains a high level of uniformity while also allowing for enough personalization to engage users and positively impact their health and well-being. This personalization should motivate patients to complete the recovery process without altering the essential activities required for recovery. Some participants were not receptive to including emotional well-being as part of the curriculum, posing a design challenge. For some, especially those who may not recognize the importance of engaging with the well-being aspects of the curriculum, offering a comprehensive orientation that emphasizes the importance of these components could improve engagement and ensure they benefit fully from all aspects of the recovery process. For others, the option to skip such content may be the only way they decide to engage with the other components of the system. Finding a balance between uniformity and personalization in the \textit{Care Curriculum} will be necessary to ensure that all patients receive effective, equitable, and comprehensive care. 

\paragraph{Including without caricaturizing}
Most participants emphasized the significance of incorporating Black culture into the \textit{Care Curriculum} to increase their engagement. This finding aligns with prior research, which found that excluding Black culture from voice-assisted health programs perpetuates inequities among Black older adults \cite{harrington_its_2022}. Integrating Black culture could involve incorporating books familiar to the Black community, featuring Black voice assistants and accents, and including music popular within the Black community. Participants expressed that such features would increase their motivation to engage with the content. A crucial consideration is that these design decisions should be made in consultation with members of the Black community to avoid perpetuating harmful stereotypes or caricatures. This finding is consistent with previous work on the importance of including cultural representation, such as Black dialects, in ways that avoid reinforcing racial stereotypes \cite{brewer2023envisioning}.

\paragraph{Norms versus social progress}
Some of the participants appeared to either align with traditional gender and social norms or their personal preferences aligned with these norms, while others did not seem to align their gender with personal preference. The \textit{Care Curriculum} introduced a digital garden allowing users to engage in physical activities with a garden theme. However, many men and a few women participants showed reluctance towards including this garden metaphor preferring sports or casino metaphors. These complex reactions and responses suggest that the \textit{Care Curriculum} should not be designed based solely on traditional gender stereotypes (e.g., assuming that men will only be interested in typically male-oriented content). It shows the importance of recognizing that people have individual preferences. These preferences may or may not align with conventional gender norms, as many individuals are non-binary, or are binary but have preferences traditionally associated with another gender. To accommodate these diverse personal interests, the \textit{Care Curriculum} should offer a range of options—
allowing users to choose what that they find personally engaging, meaningful, and relevant.

\subsection{The future of the Care Curriculum}
Our intention with this research is to eventually deploy inclusive \textit{Care Curriculum} features across care settings, from the hospital, the clinic, and into the home care setting. In this section, we first describe the initial steps we will take, as a way to gain additional context through the lens of the broader healthcare community and enact the findings and design considerations described earlier. Then, we will list design opportunities for the \textit{Care Curriculum} moving forward, including increasing cultural relevance, addressing stigma, and balancing potential good with risk of harm. 

\subsubsection{Inclusion of healthcare providers}
To fully integrate the \textit{Care Curriculum} into the broader infrastructure of recovery care, it is key to include healthcare providers in the process and understand their interactions with the system. Previous research has investigated how home health aides can use IVAs to support their work \cite{bartle2022second, bartle2023machine}. Our study contributes a defined system that healthcare providers and caregivers can work with. While the \textit{Care Curriculum} offers a structured and largely independent system, caregivers can play a role in personalization, supporting, updating, and management. 

\subsubsection{Minimum viable interface}
Our initial deployment would require at least two interfaces, one for the provider (see Figure \ref{fig:prov-int} in the Appendix), and one for the patient (see Figure \ref{fig:pat-int}, also in the Appendix). Based on initial frailty assessments and symptom status (e.g., pain and fatigue levels, etc.), the provider can select between three default options for the \textit{Care Curriculum}: for patients who are 1) ambulatory, 2) chair-, 3) or bed-bound, and can modify the curriculum according to each patient's mobility, functional ability, and learning needs. This design choice supports the need for adaptability that surfaced in our study, and future designs should increase opportunities to incorporate individual preferences for information delivery, culturally relevant content examples and personas, and incremental goal-setting as function improves. The patient interface is now stripped of any content related to media or metaphors with cultural signifiers that could push people away. It also includes a personable chat where the provider or an AI chatbot can send messages to the patient and also provide explanations to justify the relevance of content within different curriculum modules. The interface could also include dashboards or displays of information about the patient for health management allowing them to make informed personal decisions during the recovery process.

\subsubsection{Design opportunities}
As technology advances, future design opportunities abound to leverage voice interaction with AI agents to truly customize the very general health promotion recommendations. 

\paragraph{Designing for cultural relevance}
We based the original content of the \textit{Care Curriculum} on national clinical survivorship guidelines \cite{rock2022american,national2023nccn,ligibel2022exercise, shams2019operationalizing}, which are extremely broad, and little work has been done to test implementation across racial and ethnic patient groups. There are opportunities to seek out underrepresented colleagues and experts from other disciplines who not only bring needed subject matter expertise but also lived experiences from the cultural perspective we wish to design for. As we expand and refine the content of the curriculum to continually improve cultural relevancy, community experts should be hired to change generic content to one that is culturally relevant across populations, much as we ask language service providers to not just translate text verbatim but to include important cultural nuances that convey true meaning. For example, a patient with heart disease working to improve the quality of their diet may ask the IVA for information on how to adapt favorite recipes to meet clinical recommendations made by their provider, such as to decrease saturated fat or sodium content yet preserve the regional or cultural authenticity of a dish (see Figure \ref{fig:recipe} in the Appendix). 

\paragraph{Addressing stigma}
Participants felt they did not need a system like this because they considered themselves highly functional, despite some having experienced adverse health events. This finding suggests that once a patient feels better and perceives themselves as highly functional, they may no longer use the curriculum, even if they still have content to cover. This idea presents a design and healthcare challenge in terms of encouraging continued use for preventative care or using the system as a preventative measure. To mitigate this, two key areas need to be addressed: 1) convincing patients to use the system long-term for continued care and preventative purposes while in recovery, and 2) ensuring the system supports changes in ability in a way that attracts individuals who perceive themselves as completely physically healthy but still in need of the system's benefits. Therefore, the system should be designed to adapt its content and interface to accommodate the changing needs and abilities of older adults as they progress through different stages of aging and health conditions in recovery. To achieve this, the system can be integrated into a publicly accessible application that caters to individuals of all abilities, ages, and personal health goals. This inclusive approach allows for a wide range of applications. For example, individuals in knee surgery recovery, a more common operation, can be presented with initial assessment questions and provided with a tailored \textit{Care Curriculum} to address their specific needs. Similarly, people undergoing muscle recovery, which is generally less stigmatized, can also benefit from a customized \textit{Care Curriculum} to undergo that process accurately.

\paragraph{Balancing potential for good with risk of harm}
Many questions arise about how to balance technological opportunities with user agency and consent, for which much concern was expressed in this study. The technology that enables Carol to ``notice'' behaviors, which requires always-on sensors, may unlock features for early detection of clinical changes or unmet needs, but raise concerns about how users can remain aware of and in control of such data collection, as indicated in Section \ref{sec:consentfulinteractions}. Current communication with the IVA driving the \textit{Care Curriculum} is primarily initiated by the patient for this particular use case, but this limits the potential benefits of functions like reminders, sedentary behavior-initiated ecological momentary assessments, and other IVA-initiated interactions. 
Designers should ensure that the \textit{Care Curriculum} prioritizes obtaining consent from the start and offers opportunities for individuals to modify their consent as needed. For example, the IVA may ask every couple of weeks whether the user wants it to initiate reminders to complete care tasks throughout the day or if the user would rather contact the IVA for reminders (see Figure \ref{fig:reminder} in the Appendix). While this method can result in gaps in completing the \textit{Care Curriculum}, significant gaps could be addressed by the user's social community (see Figure \ref{fig:caregiver} in the Appendix) or healthcare provider during follow-up appointments. 

A way to address privacy concerns is by considering the physical location of the device with the \textit{Care Curriculum}. Placing the device in more public spaces in the home (e.g., living room versus bathroom) could help indicate the level of privacy to be expected. If the \textit{Care Curriculum} is mobile, such as via a phone app instead of via a static smart speaker, then the device can detect the different privacy levels of the different rooms and modify the content accordingly. Similarly, if it detects people around, it can behave differently than if it does not. This feature is similar to a design consideration described by \citet{cuadra2023designing}, the ``type and level of companionship should be dynamically personalized to a person’s preferences and context.'' In this case, it extends beyond companionship into all types of communication. 

An added consideration for incorporating the benefits of always-on sensing involves the IVA’s ability to analyze the user's speech patterns. This feature allows the IVA to provide voice options using learned dialects, languages, tones, and accents. Additionally, this capability enables the IVA to better understand the communication style of older adults over time, allowing them to interact with the system using their natural speech patterns. A compromise could be allowing selective disabling of more intrusive sensing capabilities while maintaining others. This approach would balance functionality with privacy concerns. Designers would need to understand how different sensing abilities interact to support older adults' recovery. Offering various sensing ``packages'' and clearly communicating their effects could appeal to users, giving them greater control over their comfort level with the system's monitoring capabilities.

Other considerations affecting user agency and consent may include user preferences about whether IVA hardware should be visible or hidden. Some of the participants alluded to the physical presence of the device as significant, both in the reminder of its presence and function, as well as the need to relocate it between rooms in the home for different activities. Other research teams have been investigating the functionality and outcomes associated with instrumented ``smart homes'' to support aging in place, where multiple sensors are invisibly active throughout the house \cite{demiris2023using, popejoy2023development, consolvo2004carenet, consolvo2004technology}. As such, it is crucial for the \textit{Care Curriculum} to consider the possibilities of the technology in conjunction with how these possibilities impact user agency and consent.

\begin{table}[H]
\caption{Design Recommendations for the \textit{Care Curriculum}}
\small
\begin{tabularx}{\textwidth}{lX}
\hline
\textbf{Category} & \textbf{Design Recommendations} \\ \hline
Content & \textbf{Recovery guidance:} Provide orientation to the recovery program; Implement introductions for each task that connect them to recovery; Incorporate the ability to interact with an AI chatbot for questioning \\ & \textbf{Curriculum personalization:} Create a modular recovery curriculum with core components and customizable modules tailored to individual needs and preferences \\ \hline
Care & \textbf{Support systems:} Develop an always-available system to complement caregivers; Incorporate opportunities for interactions with family and support system \\ \hline
Adaptability & \textbf{User customization:} Integrate with platforms storing user preferences to shape content delivery (e.g., music and book preferences) \\ 
& \textbf{Ability accommodation:} Adjust personalization to support various ability levels; Implement progressive reminders \\ 
& \textbf{Integration features:} Develop a system that continuously adapts to changing recovery stages and needs; Integrate with medical team \\ \hline
Interactions & \textbf{User consent:} Ensure continued consensual interactions \\ 
& \textbf{User engagement:} Integrate overall health management, decision-making support, and treatment adherence tools  \\
& \textbf{IVA companionship:} Accommodate diverse views on the IVA's role as a companion
\\ \hline
Systemic barriers & \textbf{Social lens:} Understand the barriers of participation for the community being designed for; Consider infrastructure gaps affected by socioeconomic status and limited access to healthcare and technology \\ \hline
Inclusion & \textbf{Health equity:} Ensure health content is scientifically and medically accurate relative to the demographic characteristics and health profile of the user \\ 
& \textbf{Communication options:} Allow users to select synthetic voices; Offer communication styles that match user preferences \\ 
 & \textbf{Stereotypes:} Avoid using demographic-based and other stereotypes \\ & \textbf{Minimal interface:} Avoid user interface features that are not critical to the user's care \\ \hline
Accountability & \textbf{Transparent liability:} Define and explain liability policies; Incorporate elements to communicate the risks to the user \\ \hline
Privacy & \textbf{Contextual intelligence:} Demonstrate spatial awareness to prevent sharing contextually inappropriate information \\ \hline
\end{tabularx}
\label{tab:recs}
\end{table}

\section{Limitations and future work}
The goal of the \textit{Care Curriculum} is to be inclusive so that all patients who have experienced serious illness feel empowered to engage during recovery. A limitation of the study is its primary focus on highly educated Black older adults. The perspective of this population may not be generalizable to all older adults of various racial, ethnic, educational, and/or socioeconomic backgrounds; therefore, expanding the scope of this research to include variations in these factors would result in more comprehensive design evidence. Future work should focus on developing a curriculum that balances cultural sensitivity, accessibility, and inclusivity, ensuring it meets the diverse needs of all patients. Additionally, the relatively small sample size of 26 participants and the qualitative nature of this study may limit its generalizability.  

Further research in this area could also include varied features in the \textit{Care Curriculum}, such as differences in the voice assistant's race, ethnicity, voice tone, and accent, as well as the individuals depicted. Investigating how perceptions change as a result of these changes could provide useful insights into user preferences and cultural factors. This study also highlighted the heightened challenges that low-income communities may face, such as difficulties in access and disparities in health infrastructure support. Future studies should involve individuals across the socioeconomic spectrum to gain a more comprehensive understanding of the challenges associated with being low-income in the context of the \textit{Care Curriculum}. 

Additional future studies could complete this research with caregivers to understand their perspective of these systems. \textit{Care Curriculum} will be deployed in collaboration with the existing recovery of patients to provide more personalized care. For those with caregivers, it is important to understand how the \textit{Care Curriculum} can motivate them to interact with it and identify features that provide mutual benefits. 

\section{Conclusion}
Our research provides a focused study into the motivations of Black older adults engaging with the \textit{Care Curriculum}. It explores their perceptions of using IVAs for enacting medical recovery at home and identifies opportunities and challenges with the system. The study highlights the importance of inclusive representation of Black culture in IVAs and the \textit{Care Curriculum}, avoiding stereotypes while ensuring cultural relevancy. Additionally, the findings emphasize the value placed by Black older adults on the \textit{Care Curriculum} accommodating all ability levels and socioeconomic statuses. The study reveals positive curiosity among Black older adults regarding cutting-edge technology, recognizing the practical opportunities of incorporating these features into products tailored to them. We enacted the lessons from this study through future design plans. Our research builds upon prior studies by identifying opportunities to integrate AI and innovative technological solutions for challenges associated with marginalization and aging in place.

\appendix
\section{APPENDIX}

Here, we include images and sketches for the future of the \textit{Care Curriculum} based on the findings from this study. 

\begin{figure}[H]
\includegraphics[width=.8\textwidth]{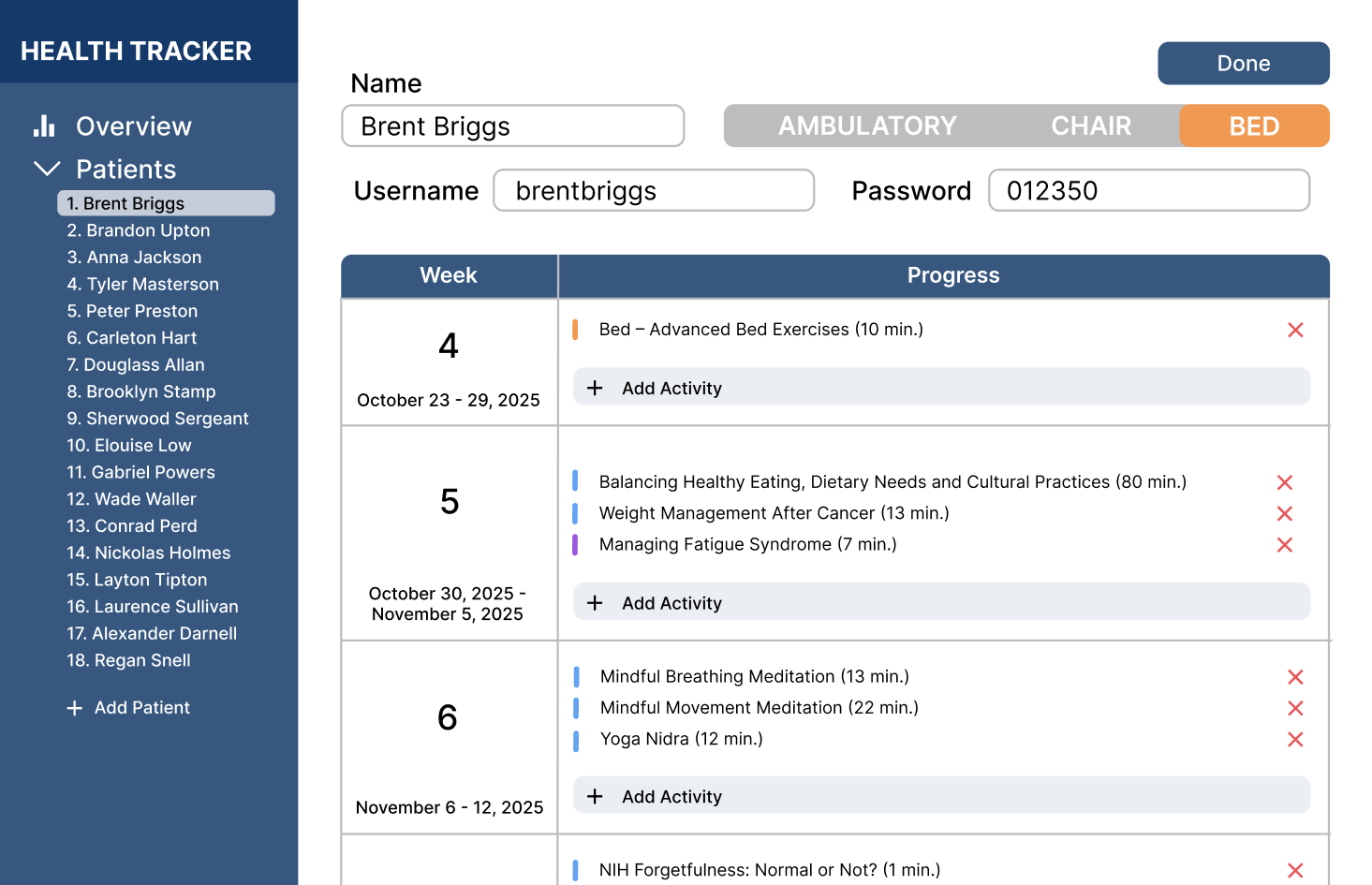}
    \caption{Medical provider-facing interface. All names are fictional.}
    \label{fig:prov-int}
\end{figure}

\begin{figure}[H]
    \includegraphics[width=.6\textwidth]{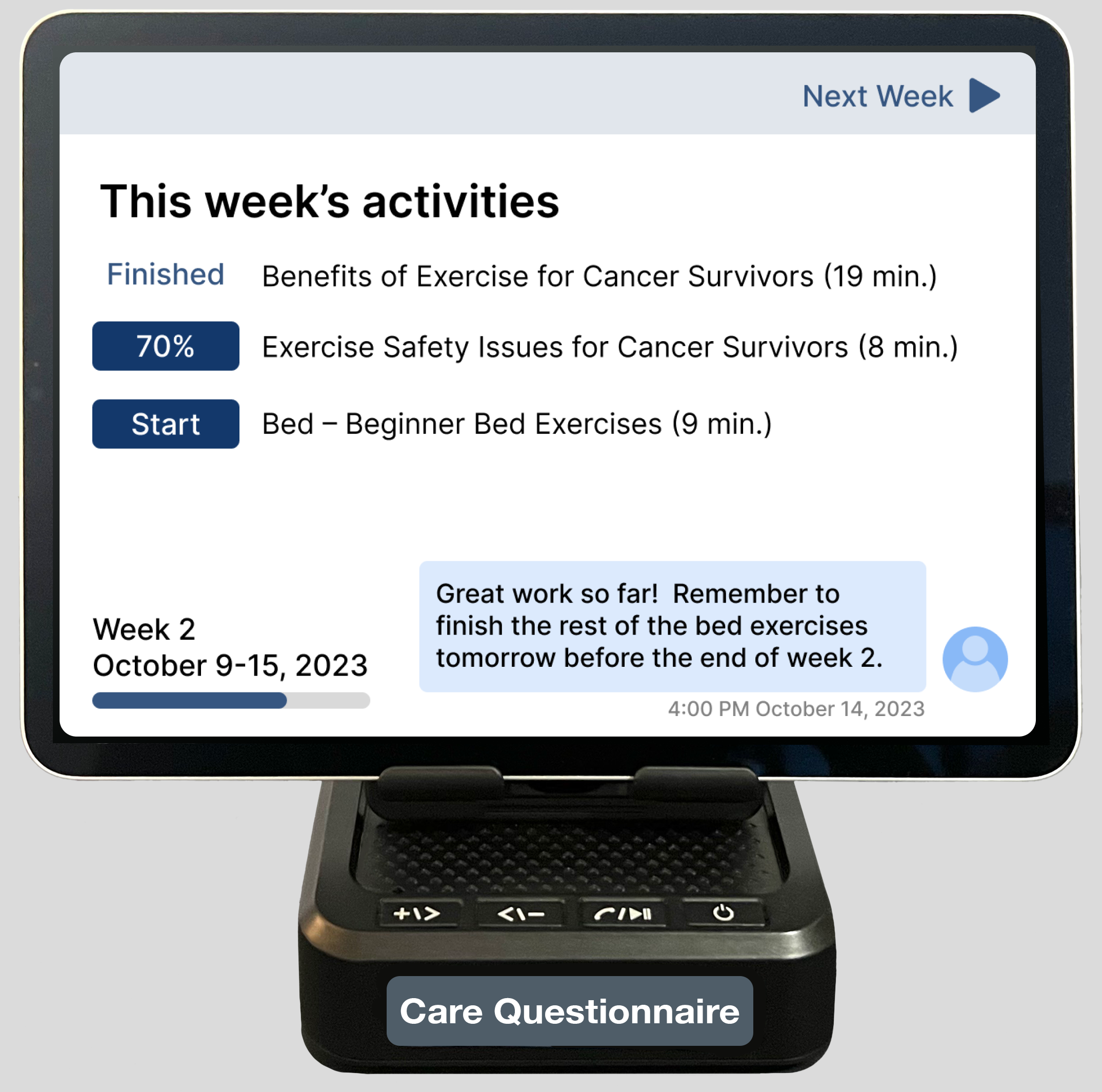}
    \caption{Patient interface, controllable by both voice and touch.}
    \label{fig:pat-int}
\end{figure}

\begin{figure}[H]
    \includegraphics[width=.6\textwidth]{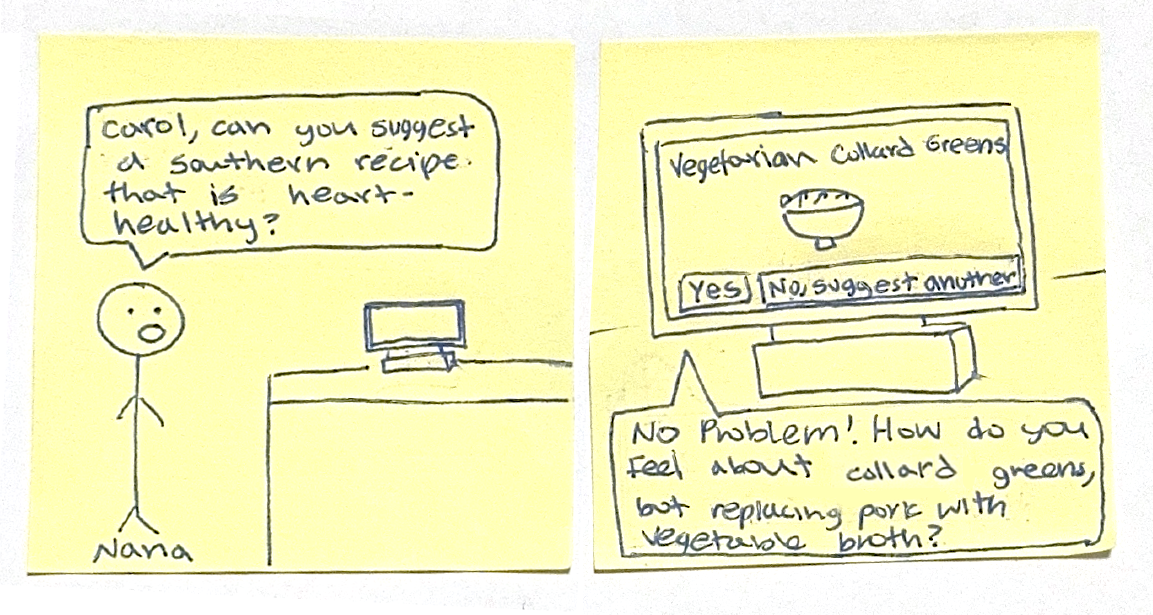}
    \caption{Recipe sketch.}
    \label{fig:recipe}
\end{figure}

\begin{figure}[H]
    \includegraphics[width=.3\textwidth]{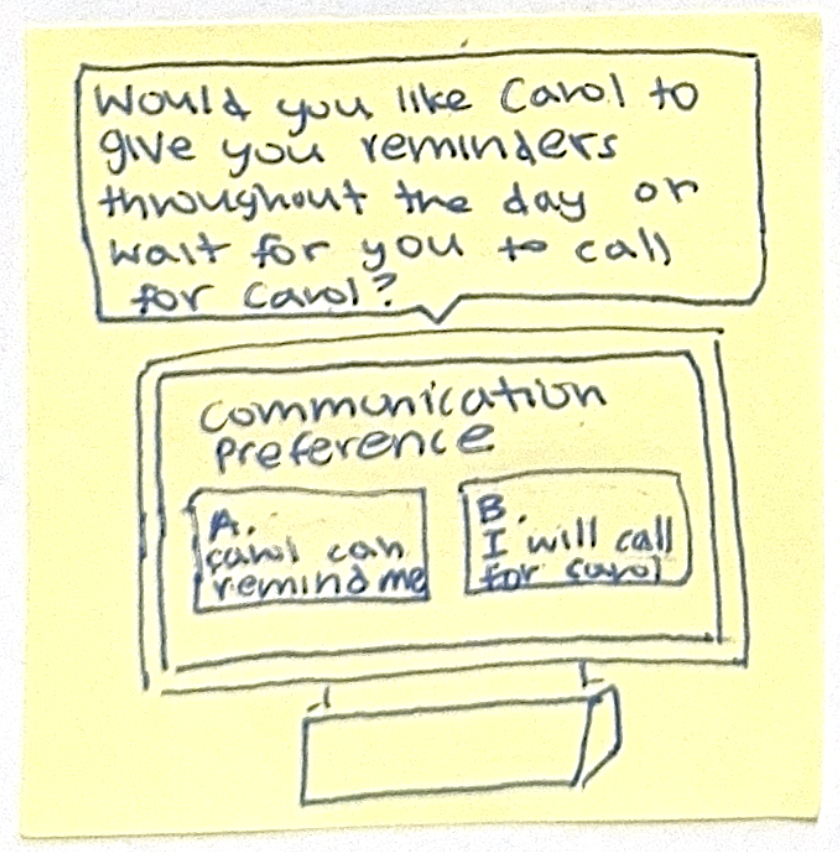}
    \caption{Reminders preference sketch.}
    \label{fig:reminder}
\end{figure}

\begin{figure}[H]
    \includegraphics[width=.6\textwidth]{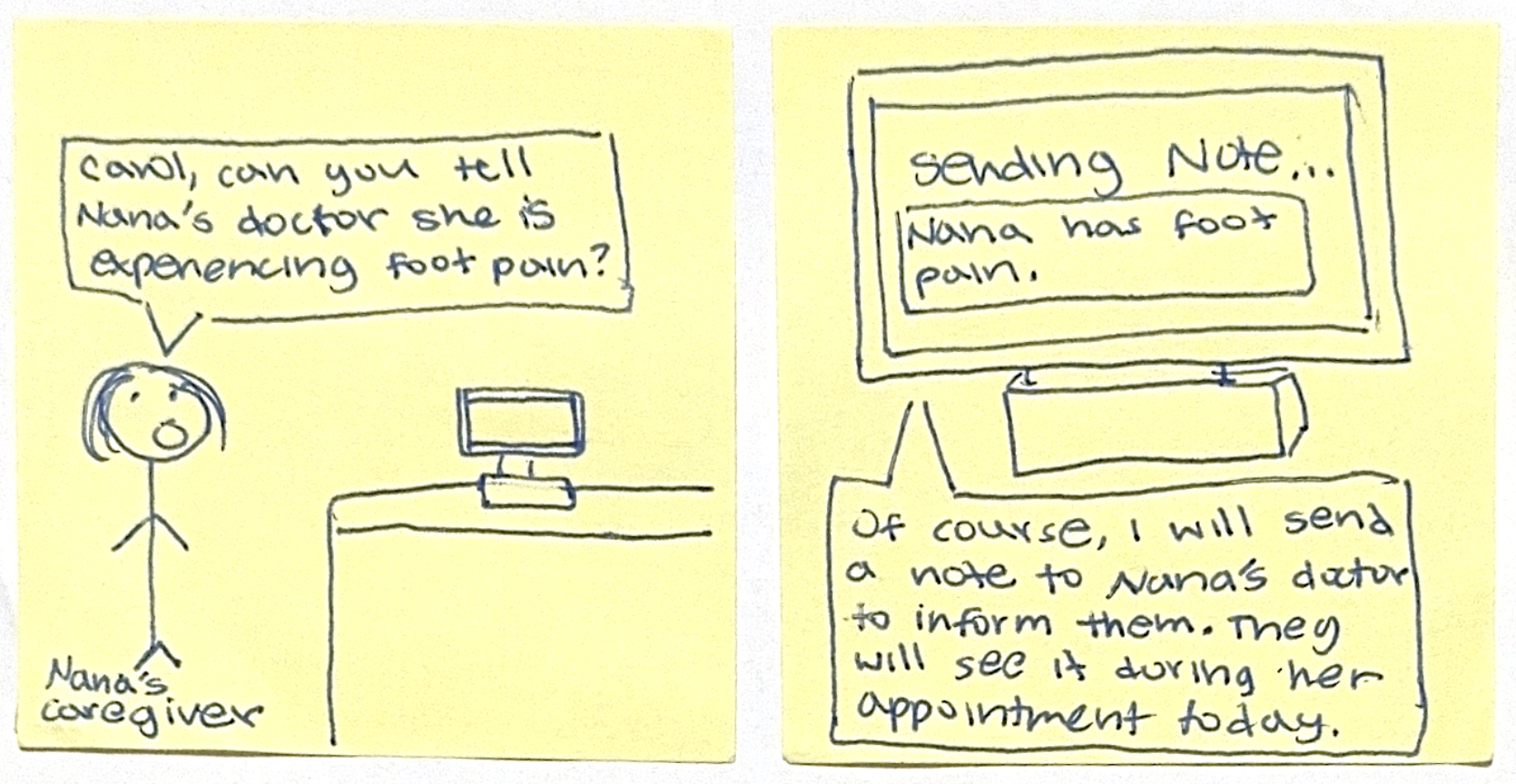}
    \caption{Caregiver communicating with healthcare provider sketch.}
    \label{fig:caregiver}
\end{figure}

\begin{acks}
We would like to express our gratitude to several individuals and organizations who contributed to this work. First, we are grateful for the participants for their role in this research. Next, special thanks to Carmen Deshon and Angelita Dubón for their constant support of this work and for helping us create the design probes. Finally, we appreciate the support from the INSPIRE CS program and the feedback from the HPDS group. This research was partially supported by Stanford HAI and a Memorial Sloan Kettering Cancer Center Support Grant/Core Grant (Grant No. P30 CA008748) funded by the National Cancer Institute, as well as through the Gordon and Betty Moore Foundation (Grant No. GBMF9048), awarded to Dr. Fessele.
\end{acks}

\bibliographystyle{ACM-Reference-Format}
\bibliography{main}

\end{document}